\documentclass[a4paper, 11pt]{article}
\usepackage{amsmath, amsthm,amssymb,mathtools}
\usepackage{kbordermatrix,blkarray}
\usepackage{tgtermes}
\usepackage{enumitem}
\usepackage[T1]{fontenc}
\usepackage{framed}

\usepackage{times}
\usepackage{graphicx}
\usepackage{xcolor}
\usepackage{subcaption}

\newcommand{\bproof}{\noindent {\bf Proof.} }
\newcommand{\eproof}{\noindent $\blacksquare$}
\newcommand{\eproofs}{\noindent $\blacksquare$\medskip}
\newcommand{\searrownew}{ \raisebox{2.5pt}{\rotatebox{15}{\resizebox{0.35cm}{!}{${\searrow}$}}}}
\newcommand{\nearrownew}{ \raisebox{5.5pt}{\rotatebox{345}{\resizebox{0.35cm}{!}{${\nearrow}$}}}}

\newtheorem{thm}{Theorem}
\newtheorem{lem}{Lemma}
\newtheorem{prop}{Proposition}
\newtheorem{cor}{Corollary}

\newtheorem{rmk}{Remark}
\newtheorem{defin}{Definition}
\newtheorem{ex}{Example}
\newtheorem{pb}{Problem}

\newcommand{\precf}{\prec^{\circ}}

\newcommand{\preceqf}{\preceq^{\circ}}
\newcommand{\precw}{\prec^{w}}

\newcommand{\preccso}{\prec^{cso}}
\newcommand{\Last}{\it Last}
\newcommand{\Prev}{\it Prev}
\newcommand{\Next}{\it Next}
\newcommand{\Front}{\it \mathcal{F}}
\renewcommand{\kbldelim}{[}
\renewcommand{\kbrdelim}{]}

\newcommand{\bp}{\begin{pb}\rm}
\newcommand{\ep}{\end{pb}}
\newcommand{\br}{\begin{rmk}\rm}
\newcommand{\er}{\end{rmk}}
\newcommand{\bdefin}{\begin{defin}\rm}
\newcommand{\edefin}{\end{defin} }
\newcommand{\bex}{\begin{ex}\rm}
\newcommand{\eex}{\end{ex}}

\newcommand{\bthm}{\begin{thm}}
\newcommand{\ethm}{\end{thm}}
\newcommand{\blem}{\begin{lem}}
\newcommand{\elem}{\end{lem}}
\newcommand{\bprop}{\begin{prop}}
\newcommand{\eprop}{\end{prop}}
\newcommand{\bcor}{\begin{cor}}
\newcommand{\ecor}{\end{cor}}

\usepackage{algorithm}
\usepackage{algorithmic}

\usepackage{xargs}                      

\setlist[itemize]{noitemsep, nolistsep}
\setlist[enumerate]{noitemsep, nolistsep}

\begin{document}
\vspace*{-1cm}
\begin{center}
{\large \sc Forced pairs in $A$-Stick graphs }
\bigskip

Irena Rusu

{\it LS2N, University of Nantes, France\\
\small Irena.Rusu@univ-nantes.fr}
\end{center}
\bigskip\bigskip

\begin{center}
\begin{minipage}[h]{13cm}
\paragraph{Abstract}
 {\small A Stick graph $G=(A\cup B, E)$ is the intersection graph of a set $A$ of horizontal segments and a set $B$ of vertical segments 
 in the plane, whose left and respectively bottom endpoints lie on the same ground line with slope $-1$. These endpoints
  are respectively called  $A$-origins and $B$-origins.
 When a total order is provided for the $A$-origins, the resulting graphs are called $A$-Stick graphs. 
 
 In this paper, we propose a characterization of the class of $A$-Stick graphs using {\em forced pairs}, which are pairs of segments
 in $B$ with the property that only one left-to-right order of their origins is possible on the ground line. We deduce a recognition algorithm for $A$-Stick graphs running 
 in  $O(|A|+|B|+|E|)$ time, thus improving the running time of $O(|A|\cdot |B|)$ of the best current algorithm. 
 We also introduce the problem of finding, for a Stick graph, a representation using segments of minimum total length. 
 The canonical order on the $A$- and $B$-origins, output by our recognition algorithm, 
 allows us to obtain
 partial results on this problem. 
}
 
 \end{minipage}
 \end{center}
 
\section{Introduction}

Defined as the intersection graphs of a set of intervals on the real line, {\em interval graphs} have been 
intensely studied, and led to the introduction of many other classes of intersection graphs. 
Various geometric objects in one or two dimensions, like segments, half-lines, arcs on a cycle, trapezoids, curves in the plane, polygons 
with corners on a cycle etc. have been used to define intersection graphs \cite{brandstadt1999graph,mckee1999topics,chaplick2018grid}, 
and a series of applications to electrical networks, nano PLA-design, computational biology, traffic control have been identified  
\cite{sinden1966topology,mckee1999topics,shrestha2011two,halldorsson2011clark,baruah2013intersection}.

In \cite{chaplick2018grid}, the authors study the relations between some existing classes of intersection graphs, chosen
for their applications and their algorithmic features, but
also introduce new intermediate graph classes. Among them, Stick graphs are defined as the intersection graphs of horizontal and vertical segments 
whose left and respectively bottom endpoints belong to a ground straight line with slope -1. Stick graphs are therefore bipartite graphs,
in which the set of horizontal (respectively vertical) segments is denoted $A$ (respectively $B$). The segments are  also named
$A$- and $B$-segments respectively. The endpoints lying
on the ground line are called $A$- or $B$-origins, depending on the type of segment, whereas the other endpoint of each
segment is called its tip (following \cite{rusu2020stick}). The origins are assumed to be ordered from left to right
on the ground line. When an order on the $A$-origins is provided, any Stick graph whose $A$-origins satisfy that order is
called an $A$-Stick graph. When both an order on the $A$-origins and an order on the $B$-origins are provided, any Stick graph 
satisfying them is called an $AB$-Stick graph. 

\begin{figure}[t!]
 \centering
 \vspace*{-2.5cm}
 \includegraphics[width=0.85\textwidth]{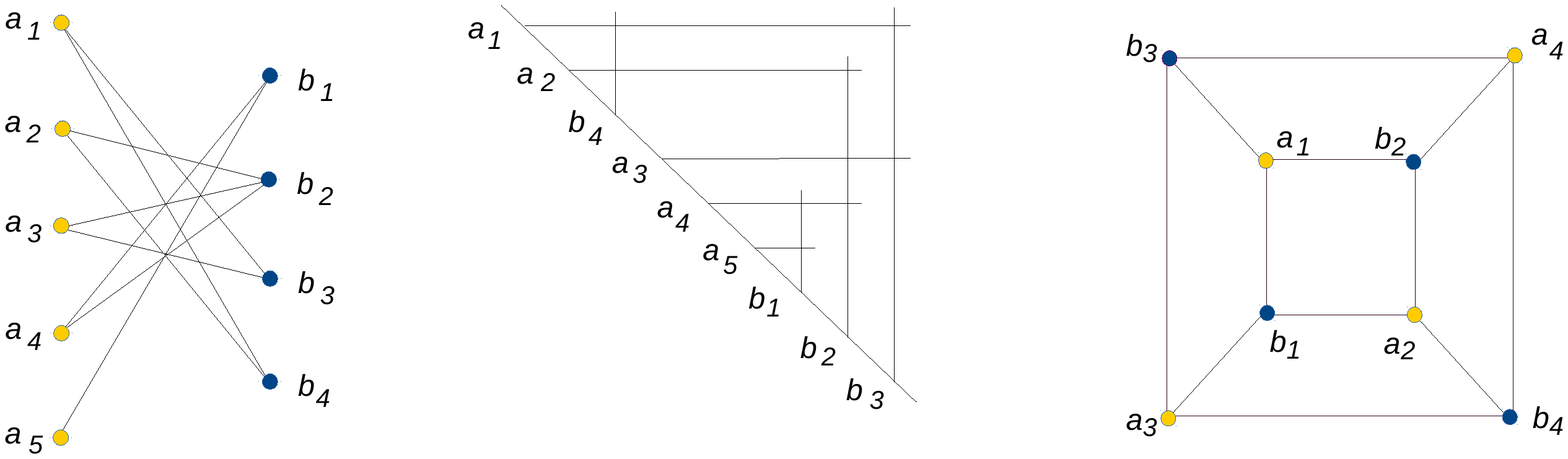}
 \vspace*{-5cm}
 \caption{\small A Stick graph and one of its Stick representations. A graph that is not Stick \cite{luca2018recognition2}.}
  \label{fig:exStick}
\end{figure}

The problem of recognizing Stick graphs is denoted {\sc STICK}. Given a graph $G=(A\cup B,E)$, it requires to test whether $G$ has
a {\em Stick representation}, consisting in an horizontal segment $A_i$ for each vertex $a_i\in A$, a vertical segment $B_i$ for each 
vertex $b_i\in B$ and an order on their origins such that $A_i$ intersects $B_j$ if and only if $a_ib_j\in E$. 
The origin of the segment $A_i$ ($B_j$) is denoted, as the corresponding vertex, by $a_i$ ($b_j$). An example of a Stick graph and one of its Stick representations is given in Figure \ref{fig:exStick},
together with a graph that is not Stick.
When $A$-Stick (respectively $AB$-Stick) graphs are concerned, the problem is usually 
denoted {\sc STICK$_A$} (respectively {\sc STICK$_{AB}$}), and we keep this notation. We formulate {\sc STICK$_A$}, the problem we
are more particularly interested in, more precisely below: 
\newpage

\noindent{\bf\sc STICK$_A$}

\noindent {\bf Input:} A bipartite graph $G=(A\cup B,E)$ and a total order $a_1, a_2, \ldots, a_{|A|}$ on the elements in $A$.

\noindent {\bf Question}: Is there an $A$-Stick representation for $G$, {\em i.e.} a Stick representation such that the order of the $A$-origins on the ground line from left to
right is $a_1,a_2, \ldots, a_{|A|}$?
\bigskip

An  $AB$-Stick representation is defined similarly. The problem {\sc STICK} is open, but
{\sc STICK$_A$} and {\sc STICK$_{AB}$} are polynomial \cite{luca2018recognition2}. The best algorithms for them have running times in
$O(|A|\cdot |B|)$ and respectively in $O(|A|+|B|+|E|)$ \cite{chaplick2019recognizing2}. 

In this paper, we provide a characterization of $A$-Stick graphs using {\em forced pairs}, which are pairs of segments in $B$
whose origins have a fixed order on the ground line, decided by the structure of $G$. We show that five simple rules allow us to
identify a set of forced pairs whose acyclicity is necessary and sufficient to decide that $G$ is an $A$-Stick graph. 
This new approach leds to an algorithm for {\sc STICK$_A$} running in $O(|A|+|B|+|E|)$ time, that improves the $O(|A|\cdot|B|)$ 
running time of the best current algorithm 
\cite{chaplick2019recognizing2}. Our algorithm builds a so-called {\em canonical} order on $A\cup B$, which 
is then used to obtain partial results on a new problem we propose. 
The length of the segments in a Stick representation is an important parameter, and was considered in \cite{chaplick2019recognizing2}
as a supplementary input to the problem of recognizing Stick graphs. More precisely, the problem asking whether a given graph has a Stick 
representation where the length of each segment is provided has been proved $NP$-complete in the three contexts: STICK, STICK$_A$ 
and STICK$_{AB}$.
Here, we consider the problem of finding, for a Stick graph, a Stick representation in which the total length of the segments is
minimum. We show that the problem is easily solved using canonical orders when reduced to $AB$-Stick graphs, and give a partial result 
for the case of $A$-Stick graphs. 

\section{Definitions, notations and preliminary results}\label{sect:Intro}

We use directed acyclic graphs (or DAG) as well as undirected graphs, for which the notations are classical. The bipartite undirected 
graph  we consider in the whole paper is denoted $G=(A\cup B, E)$. The vertices adjacent to a vertex $c\in A\cup B$ are called its
{\em neighbors} and form the {\em neighborhood} $N(c)$ of $c$ in $G$.

Given a DAG $H$, we denote $V(H)$ (respectively $E(H)$) its set of vertices (respectively of arcs). The subgraph induced 
in $H$ by a subset $V'$ of $V(H)$ is denoted $H[V']$ and is called an {\em induced sDAG} of $H$. 
The difference $H$-$F$ of a DAG $H$ and its induced sDAG $F$ is the sDAG of $H$ induced
by $V(H)\setminus V(F)$. Furthermore, we define the {\em in-neighborhood} of $x\in V(H)$, and denote it $N^-(x)$, to be the set of vertices $z$ such 
that $zx\in E(H)$. These vertices $z$ are the {\em in-neighbors} of $x$. Similarly, the {\em out-neighbors} $z$ of $x$ satisfy $xz\in E(H)$
and form the {\em out-neighborhood}  $N^+(x)$ of $x$. When $N^-(x)=\emptyset$, $x$ is called a {\em source} of $H$,
whereas when $N^+(x)=\emptyset$, $x$ is a {\em target} of $H$. If a directed path exists in $H$ from $x$ to $y$, 
then we say that $x$ is a {\em predecessor} of $y$, that $y$ is a {\em successor} of $x$ or that $x$ {\em precedes} $y$.
When the path is chordless and has exactly two edges,  $x$ is called a {\em depth-2 predecessor} of $y$, and $y$ is called a 
{\em depth-2 successor} of $x$. 

By convention, the
arcs of a DAG are considered to be directed from left to right, so that $x$ precedes $y$ if and only if $y$ is to the right of $x$.
\nopagebreak

Before using DAGs, we focus on the constraints induced by the order provided on the $A$-origins in an $A$-Stick graphs. They are
investigated in the next section, and use the result given below, proved in \cite{luca2018recognition2}.
Given an order $\sigma_A$ on $A$ and an order $\sigma_B$ on $B$, the {\em ordered adjacency matrix} 
$M$ of $G$ is the adjacency matrix of $G$ whose rows (respectively columns) are the $A$-origins (respectively $B$-origins)
in increasing order according to $\sigma_A$ (respectively $\sigma_B$) from top to bottom (respectively from left to right).  
A $*$ indicates a value that can be either 0 or 1. An {\em ordered submatrix} of $M$ is any submatrix made of the elements at the 
intersection of a set of rows and a set of columns of $M$, following the same order of rows and columns as in $M$.

\bthm \cite{luca2018recognition2}
An instance of Stick$_{AB}$  has a solution if and only if the ordered adjacency matrix $M$ of $G$ has no 
ordered submatrix of the following form:

\[
  \mathbf{P_1} = 
    \kbordermatrix{ & b_p & b_q & b_r \cr
      a_i & * & 1 & * \cr
      a_j & * & 0 & 1 \cr
      a_k & 1 & * & * } \qquad
  \mathbf{P_2} = 
    \kbordermatrix{ & b_p & b_q\\
      a_i & 1 & * \\
      a_j & 0 & 1  \\
      a_k & 1 & *  } \qquad
  \mathbf{P_3} = 
    \kbordermatrix{ & b_p & b_q & b_r \\
        a_i & * & 1 & * \\
        a_j & 1 & 0 & 1} \qquad
\]

\label{thm:deLuca}
\ethm

The organization of the paper is as follows. In Section \ref{sec:prelim} we formulate the constraints imposed by the order on $A$
as a set of forced pairs of origins, for which only one left-to-right order is possible on the ground line. We prove that these
forced pairs are sufficient to decide whether a given bipartite graph is an $A$-Stick graph, and we show that they may
be identified using five rules, applied in a precise order. In Section \ref{sec:linalg} we describe the algorithm, prove its
correctness and give all the details of its implementation in linear time. In Section~\ref{sec:length} we introduce the minimum length 
Stick representations, the corresponding problem and give our results. Section~\ref{sec:conclusion} is the conclusion. 

In all the paper, we assume that $G$ has no isolated vertices in $B$, as these vertices
are not decisive for the existence or not of an $A$-Stick representation. 

\section{Forced pairs in $A$-Stick graphs}\label{sec:prelim}

We assume that the $A$-origins are denoted and ordered as $a_1,a_2, \ldots, a_{|A|}$, from left to right.

\subsection{Definition and characterizations}

In an $A$-Stick graph $G$, we say that a couple $(c_h,c_j)$ of origins, in this precise order, is a {\em forced pair} 
if $c_h$ must be to the left of $c_j$ in every $A$-Stick representation. 
Then every pair $(a_i,a_j)$ is a forced pair, since the order in $A$ is given, but other pairs may be forced.

For each $B$-origin $b_j$, denote $k_j=\max\{k\,|\,  a_kb_j\in E\}$ and $1_j=\min\{i\, |\, a_ib_j\in E\}$. Below, we define five rules, 
that we call {\em forcing rules}, allowing us to identify, starting with the order on $A$ only, a set of forced pairs. 
In order to distinguish between the forced pairs   $(c_i,c_j)$ obtained by 
these rules and the other forced pairs, the former ones are also denoted as $c_i\precf c_j$. 

\bprop
Let $G$ be an $A$-Stick graph, and consider the following rules: 

\noindent\begin{itemize}[leftmargin=0.6cm]
\item[] {\makebox[4.3cm]{\bf Order on $A$ (O):\hfill} If $a_i, a_j\in A$ and $i<j$ then $a_i\precf a_j$.}

\item[] {\makebox[4.3cm]{\bf Adjacency (A):\hfill} If $a_t\in A$ and $b_j\in B$ such that $a_tb_j\in E$ then $a_t\precf b_j$.}

\item[] {\makebox[4.3cm]{\bf True Betweenness (TB):\hfill} \begin{minipage}[t]{11.5cm} If  $a_s, a_t\in A$ and $b_h, b_j \in B$ 
such that $a_s\precf a_t\precf a_{k_j}$, $a_sb_j\in E,$ $a_tb_j\not\in E,$ and $a_tb_h\in E$, then $b_h\precf b_j$.\end{minipage}}
 
\item[] {\makebox[4.3cm]{\bf False betweenness (FB):\hfill} \begin{minipage}[t]{11.5cm} If  $a_t\in A$ and $b_w, b_h, b_j \in B$ 
such that $b_w\precf b_j$, $a_{k_j}\precf a_t, a_tb_w\in E$ and  $a_tb_h\in E$, 
then $b_h\precf b_j$.\end{minipage}}
 
\item[] {\makebox[4.3cm]{\bf Transitivity (T):\hfill} If $c_p, c_q, c_r\in A\cup B$ such that $c_p\precf c_q\precf c_r$, then $c_p\precf c_r$.}
\end{itemize}
Then each pair $(c_i,c_j)$ such that $c_i\precf c_j$  
is a forced pair.
\label{prop:cond}
\eprop

\begin{figure}[t]
 \centering
 \vspace*{-2.5cm}
 \includegraphics[width=0.4\textwidth,angle=270]{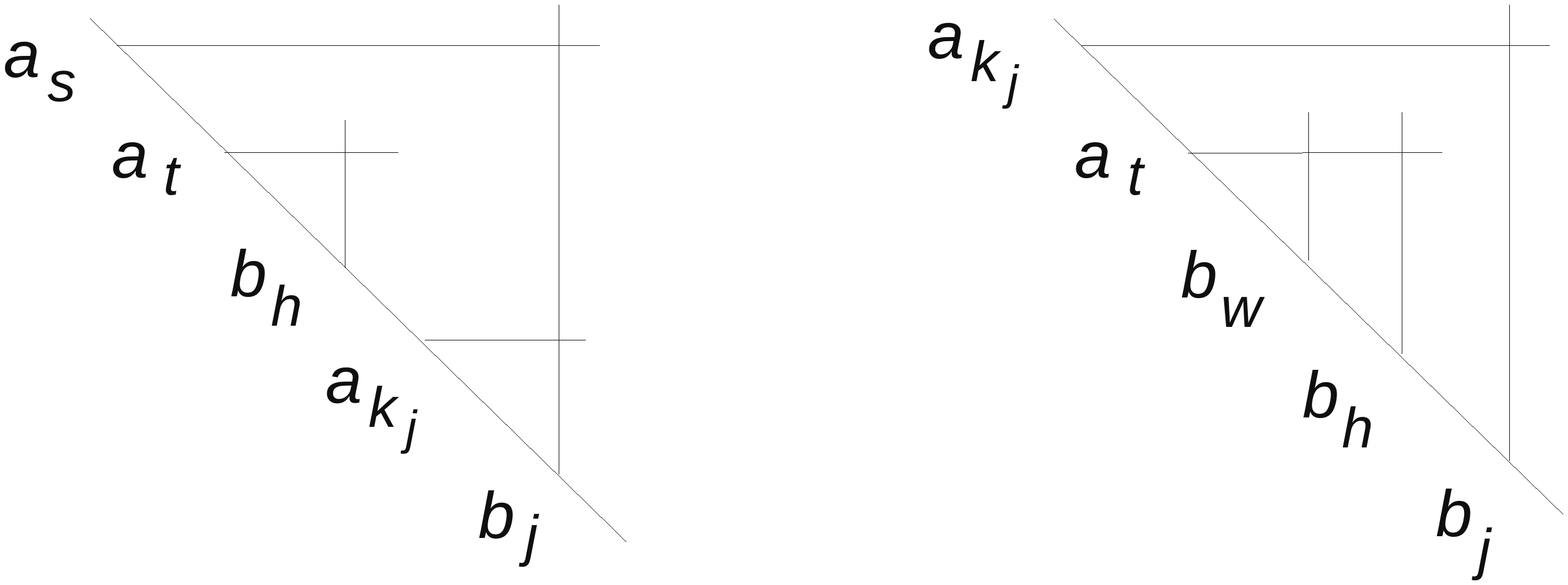}
 \vspace*{-1cm}
 \caption{\small Rules TB (left) and FB (right).}
  \label{fig:rules}
\end{figure}

Let us call a {\em forcing sequence} for $c_i\precf c_j$ each sequence
of rules that, starting with the order on $A$, finally deduces $c_i\precf c_j$. The last rule in the sequence thus  
concludes that $c_i\precf c_j$ using (O), (A), (TB), (FB) or (T).  When a rule of type (TB) or (FB) is applied, we 
call a {\em support} of $b_h$ with respect to $b_j$ the $A$-origin $a_t$  used in this rule.  The {\em length} of a forcing
sequence is the number of rules it contains. A {\em shortest} forcing sequence
for $c_i\precf c_j$ is a forcing sequence for $c_i\precf c_j$ with minimum length. 
\bigskip

\bproof ({\bf of Proposition \ref{prop:cond}}).  Let $\prec$ be an $A$-Stick representation of $G$, and let
$S$ be a shortest forcing sequence deciding that $c_i\precf c_j$. When the length of the forcing sequence
is equal to 1, only rules (O) and (A) may be applied, and the pair $(c_i,c_j)$ is a forced pair by the
order on $A$ and the definition of a Stick graph.

Assume now that  $c_{i'}\precf c_{j'}$ defines a forced pair for all pairs $(c_{i'},c_{j'})$ whose shortest
forcing sequence has length less than $n$. Then $c_{i'}\prec c_{j'}$ for all these pairs.
Let $c_i\precf c_j$ be obtained using a shortest forcing sequence
of length $n$. The last rule is (TB), (FB) or (T), and the conclusion that $(c_i,c_j)$ is a forced pair 
is immediate in case of (T), by induction hypothesis.

In order to prove (TB), assume by contradiction that $b_j\prec b_h$ is possible.
Then, with the notations $a_s, a_t$ used in the rule (TB), the adjacency matrix of $G$ ordered according to the order 
$\prec$ on $A$ and  the order $\prec$ on $B$ contains the ordered submatrix $P_2$ from Theorem \ref{thm:deLuca}.
This submatrix is given by $a_s, a_t, a_{k_j}, b_j$ and $b_h$. Then $\prec$ is not a Stick representation of $G$, a contradiction.  

In order to prove (FB), assume again by contradiction that $b_j\prec b_h$ is possible. We have $a_tb_j\not\in E$ 
(due to $a_{k_j}\precf a_t$)  and thus $a_{k_j}, a_t, b_w, b_j, b_h$, where $b_w$ is to the left of $b_j$ by inductive hypothesis, 
form the submatrix $P_3$ from Theorem \ref{thm:deLuca}, a contradiction.
\eproofs

\br
We may equivalently write rule (TB) as follows: if $a_tb_h\in E$ holds and there exist two successive neighbors $a_s, a_v$ of 
$b_j$ with respect to the order on $A$ such that $s<v<t$ (implying thus that $a_tb_j\not\in E$),  then $b_h\precf b_j$.
\label{rem:successive}
\er

\br 
No forcing rule allows to deduce $b_j\precf a_h$, for some $b_j\in B$ and $a_h\in A$. Thus in the transitivity rule (T) 
the $A$-origins are either absent, or present on the leftmost places among $c_p, c_q, c_r$ in this order.
\label{rem:T}
\er

Our aim is to show (Theorem \ref{thm:main}) that the forced pairs deduced from these rules are sufficient to determine the
nature ($A$-Stick or not) of a given bipartite graph $G$. We thus give several results concerning these pairs, before 
proving the main theorem. 

\bprop
The pair $(a_t,b_j)$ satisfies $a_t\precf b_j$ if and only if:
\begin{enumerate}
\item[(a)] either $a_t$ is a neighbor of $b_j$, or

\item[(b)] $a_t$ is a neighbor of $b_q$ such that $b_q\precf b_j$, or

\item[(c)] $t<r$ and $a_r\precf b_j$.
\end{enumerate}
\label{prop:situationsonA}
\eprop
\newpage

\bproof
The proof is by induction on the length $l_{tj}$ of a shortest forcing sequence for $a_t\precf b_j$. When
$l_{tj}=1$, only rule (A) may be applied, and (a) holds. We now prove the proposition for $l_{tj}>1$, assuming
it is true for all $l_{t'j'}<l_{tj}$.

Since $l_{tj}>1$ and the last rule of a shortest forcing sequence for $a_t\precf b_j$ can be either (A) or (T),
we deduce that the last rule is necessarily (T). Then, using Remark \ref{rem:T}, either $a_t\precf a_r\precf b_j$ 
or $a_t\precf b_u\precf b_j$ holds, with appropriate $a_r$ and $b_u$. The former case implies affirmation (c).
In the latter case, $l_{tu}<l_{tj}$
and by inductive hypothesis we deduce that for $a_t\precf b_{u}$
either (a), or (b) or (c) holds. Case (a) for $a_t\precf b_{u}$ implies case (b) for $a_t\precf b_j$. Case (b) for  
$a_t\precf b_{u}$ implies the existence of $b_p$ such that $a_t$ is a neighbor of $b_p$ and $b_p\precf b_u$;
this implies $b_p\precf b_j$, and thus case (b) holds for $a_t\precf b_j$ using $b_p$. Finally,
case (c) for $a_t\precf b_{u}$ implies case (c) for $a_t\precf b_j$.
\eproofs

True betweenness (TB) allows us to discover a new forced pair $(b_h,b_j)$ using the order on $A$ only. 
Then $b_h$ is called a {\em basis} of $b_j$. False betweenness (FB) uses an existing forced pair $(b_w,b_j)$ 
to deduce another forced pair $(b_h, b_j)$. Then we call $b_h$ a {\em child} of $b_w$. We also
call a {\em descendant} of $b_w$  any $B$-origin $b_u$ for which there exists a sequence of $B$-origins $b_{u_1}=b_w, b_{u_2}, 
\ldots, b_{u_g}=b_u$ such that $b_{u_i}$ is a child of $b_{u_{i-1}}$ for each $i\in\{2, 3, \ldots, g\}$.

\bprop
The pair $(b_h,b_j)$ satisfies  $b_h\precf b_j$ if and only if there exists a shortest forcing sequence $S$ for $b_h\precf b_j$
showing that $b_h$ is:
\begin{enumerate}
\item[(1)] either a basis of $b_j$,   or
\item[(2)] a descendant of a basis of $b_j$,  or
\item[(3)] a basis of a $B$-origin $b_q$ with $b_q\precf b_j$,  or
\item[(4)] a descendant of a basis of a $B$-origin $b_q$ with $b_q\precf b_j$.
\end{enumerate}
Cases (1), (2), (3) and (4) occur respectively when the last rule in the sequence $S$ is (TB); (FB); (FB) or (T); (FB) or (T).
\label{prop:3situations}
\eprop

\bproof
The backward direction follows by the definitions of a basis, of a descendant and of rule (T).

We now consider the forward direction. Let $n_{hj}$ be the length of a shortest forcing sequence for $b_h\precf b_j$. 
We use induction on $n_{hj}$ to show that there exists a shortest forcing sequence that yields one of the cases (1), (2), (3) or (4). 

When $n_{hj}=1$, only (TB) can be applied to deduce $b_h\precf b_j$, and then $b_h$ is a basis of $b_j$. Case (1) applies.
When $n_{hj}>1$ and the affirmation is true for smaller values $n_{h'j'}$, the last rule in any shortest forcing sequence 
for $b_h\precf b_j$  is either 
(FB) or (T). 
\bigskip

\noindent{\it Case (I): the last rule is (FB).}

Consider a shortest forcing sequence $S$ for $b_h\precf b_j$, whose last rule is (FB). Let $a_t$, with $a_{k_j}\precf a_t$,  be the support of $b_h$ with 
respect to $b_j$, and let $b_w$ with $b_w\precf b_j$  and $a_tb_w\in E$ be the other $B$-origin used by the rule.  
Then, the sequence of all rules in $S$ except the last one is a shortest forcing sequence for $b_w\precf b_j$ 
(otherwise a shorter forcing sequence would exist for $b_h\precf b_j$).
Thus $n_{wj}=n_{hj}-1$ and the inductive hypothesis may be applied for $b_w$ and a shortest forcing sequence $S_{wj}$
for $b_w\precf b_j$. Therefore $b_w$ satisfies one of the cases
(1), (2), (3), (4) with respect to $b_j$.  
Since $a_tb_w\in E$, the cases (1) and (2) for $b_w$ with respect to $b_j$ imply case (2) holds for $b_h$ with respect 
to $b_j$. Indeed, $b_h$ is the child either of a basis or of a descendant of a basis  of $b_j$.
In the cases (3) and (4) for $b_w$ with respect to $b_j$, the same holds but for $b_q$ instead of $b_j$, so 
case (4) holds for $b_h$. The forced pairs used in each case are results of the forcing sequence $S'$
given by  $S_{wj}$ followed by the final rule (FB) used in $S$. Thus $S'$ is a shortest forcing sequence for $b_h\precf b_j$
and we are done.
\newpage

\noindent{\it Case (II): the last rule is (T).}

Let $S$ be a shortest forcing sequence for $b_h\precf b_j$ whose last rule is (T). 
Then there exists $b_{q_0}$ such that the rule (T) is finally applied to $b_h\precf b_{q_0}\precf b_j$ (see Remark \ref{rem:T}). 
If $b_h\precf b_{q_0}$ is also obtained by (T) using the sequence $S$, then let $b_{q_1}$ be such that (T) is applied to 
$b_h\precf b_{q_1}\precf b_{q_0}$, and so on until we obtain $b_h\precf b_{q_m}\precf \ldots \precf b_{q_1}\precf b_q$,
and $b_h\precf b_{q_m}$ is not obtained by (T). Without loss of generality, we assume $S$ is chosen to be the shortest forcing
sequence for $b_h\precf b_j$ with last rule (T) such that $m$ is maximum.

We will show that sequence $S$ identifies $b_h$ as a basis or a descendant of a basis of $b_{q_m}$, thus showing that 
case (3) or (4) holds. (Note that using the inductive hypothesis for $b_h\precf q_m$ may not be safe since the modifications
of $S$ in order to introduce the rules of the sequence obtained by inductive hypothesis do not guarantee a resulting {\em shortest}
forcing sequence.)

Consider $b_h\precf b_{q_m}$, which cannot be obtained by (T), by the choice of $q_m$.  If $b_h\precf b_{q_m}$  
is deduced by the rule (TB), then $b_h$ is a basis of $b_{q_m}$ and $b_{q_m}\precf b_j$ such that both these affirmations are deduced from $S$. Then case (3) holds for $b_h$
and the sequence $S$. If $b_h\precf b_{q_m}$ is deduced by the rule (FB), then there exists $b_{x_1}$ such that 
$b_{x_1}\precf b_{q_m}$, deduced by $S$, and $b_h$ is a child of $b_{x_1}$. This reasoning may be repeated for 
$b_{x_1}\precf b_{q_m}$, if we are able to show that $b_{x_1}\precf b_{q_m}$ cannot be made by (T).

Assume by contradiction that $b_{x_1}\precf b_{q_m}$ is deduced by (T) using an intermediate $B$-origin $b_r$. 
Then $S$ allows to deduce $b_{x_1}\precf b_r$, $b_r\precf b_{q_m}$, and then uses them to deduce 
$b_{x_1}\prec b_{q_m}$ (using (T)) followed by $b_h\precf b_{q_m}$
(using $b_{x_1}$ and (FB)). But then the sequence $S'$ obtained from $S$ by replacing the two last deductions by
$b_h\precf b_{r}$ (using $b_{x_1}$ and $FB$) and  $b_h\precf b_{q_m}$ (using (T) on $b_h\precf b_{r}\precf b_{q_m}$) has
the same number of rules as $S$, thus is a shortest forcing sequence, but has larger $m$, a contradiction with the choice of
$S$. Thus (T) cannot be applied to deduce that $b_{x_1}\precf b_{q_m}$.

In general, assume by inductive hypothesis that $b_{x_1}, b_{x_2}, \ldots, b_{x_c}$ have been found such that
$b_{x_i}$ is a child of $b_{x_{i+1}}$ for $1\leq i\leq c-1$, and thus $b_{x_i}\precf b_{q_m}$ is obtained by (FB)
for $1\leq i\leq c$. As above, we show that $b_{x_c}\precf b_{q_m}$ cannot be obtained by (T), by postponing the use of the
rule (T) until it may be applied to $b_h, b_r$ and $b_{q_m}$, yielding a contradiction. Thus $b_{x_c}\precf b_{q_m}$
is either obtained by (TB), in which case (4) holds for $b_h$, or by (FB), in which case the same reasoning holds for the $B$-origin 
$b_{x_{c+1}}$ used in (FB). The origins $b_{x_i}$ are distinct (otherwise a shorter forcing sequence would exist), thus
the construction stops and then case (4) holds for $b_h$. All the deductions are due to rules from $S$, which has been
appropriately chosen.
\eproofs

Proposition \ref{prop:3situations} and an inductive reasoning on each $b_q$ encountered in cases (3) and (4) allow us to deduce that:

\bcor
If $b_h\precf b_j$ then there exists 
a sequence $q_1, q_2, \ldots, q_p$ of distinct integers such that $q_1=h, q_p=j$ and $b_{q_{r}}$ is either a basis or a descendant
of a basis of $b_{q_{r+1}}$, for $1\leq r\leq p-1$.
\label{cor:inductive12}
\ecor

\subsection{Main theorems}

In this subsection, we show that the forced pairs $b_h\precf b_j$ deduced from an order on $A$ provide a characterization of
$A$-Stick graphs, and also that we may limit, when we compute these forced pairs, to forcing sequences with increasing supports.

We start with two preliminary results.

\bprop
Let $b_h\precf b_j$ such that $b_h$ is a descendant of the basis $b_v$ of $b_j$, and let $a_t$ be the support of $b_h$ with respect
to $b_j$ in the last rule of the corresponding forcing sequence.
Then there exists a sequence  

$$a_{u_0}\precf a_{k_j}\precf a_{u_1}\precf b_{u_1}=b_v\precf a_{u_2}\precf b_{u_2}\precf a_{u_3}\precf \ldots a_{u_{g-1}}=a_t\precf b_{u_{g-1}}$$

\noindent such that, with the notation $b_{u_g}=b_h$, $b_{u_i}$ is a child of $b_{u_{i-1}}$ with support $a_{u_{i-1}}$, for each $2\leq i\leq g$. 
Moreover, $a_{u_0}, a_{k_j},a_1\in N(b_v)$,  $a_{u_{i-1}}, a_{u_i}\in N(b_{u_i})$ for $1\leq i\leq g-1$, and $a_{u_{g-1}}\in N(b_h)$. 
\label{prop:status2}
\eprop

\bproof
By the definition of a descendant of a basis, there is a sequence of $B$-origins 
$b_{u_1}=b_v,  b_{u_2},  \ldots,  b_{u_{g-1}},$ $b_{u_g}=b_h$
such that  $b_{u_i}$ is a child of $b_{u_{i-1}}$ for $2\leq i\leq g$, and the last rule uses the support $a_t$. 
Assume this sequence is chosen to be the shortest one with these properties. Let $a_{u_{i-1}}$ be the support of $b_{u_i}$,
for $1\leq i\leq g$. Then all the affirmations concerning the neighborhoods of $N(b_{u_{i}})$, $1\leq i\leq g$ hold, 
by the definitions of a child and of a support.
Moreover $a_{u_0}$ is the support of $b_v$ with respect to $b_j$ (rule (TB)) and $a_{u_1}$ is the support of $b_{u_2}$ (rule (FB)). 
We deduce that $a_{u_0}\precf a_{k_j}\precf a_{u_1}\prec b_{v}$ since $b_v$ is a basis of $b_j$ and  
$b_{u_2}$ is a child of $b_v$.

To show that $b_{u_{i-1}}\precf a_{u_i}\precf b_{u_i}$ for $2\leq i\leq g-1$, we use induction on $i$. 
The proof for the initial case $i=2$ (assuming $a_{u_1}\prec b_{v}$ holds) is similar to the general case
(assuming $b_{u_{j-1}}\precf a_{u_j}\precf b_{u_j}$ for all $2\leq j<i$), so that we give it only once, for both cases.
The affirmation  $a_{u_i}\precf b_{u_i}$ immediately results from rule (A), since $a_{u_i}b_{u_i}\in E$. Assume by contradiction that $a_{u_i}\prec b_{u_{i-1}}$ 
is possible in some $A$-Stick representation $\prec$, and let $l$ be such that  $a_{u_i}\prec b_{u_{l}}$ but  
$a_{u_i}\not\prec b_{u_{l-1}}$. Then $a_{u_i}b_{u_{l}}\in E$ since $b_{u_l}$ has a neighbor,
namely $a_{u_{l-1}}$, to the left of $a_{u_i}$ and $a_{u_i}$ has a neighbor, namely $b_{u_i}$, to the right of
$b_{u_l}$. But then $b_{u_{i+1}}$, which is already a child of $b_{u_i}$, is also a child of $b_{u_l}$ and therefore
the sequence of $B$-origins chosen at the beginning of the proof is not as short as possible.
Thus $b_{u_{i-1}}\precf a_{u_i}$. 
\eproofs

Let $m_j=\max\{t\, |\, a_t\precf b_j\}$. 
It is sometimes convenient to replace the rule (FB) with the rule (FB') below, which is equivalent in terms of
forced pairs that can be deduced. As before, $a_t$ is called the {\em support} of $b_h$ with respect to $b_j$.
Let $R=\{(O),(A),(TB),(FB),(T)\}$ and  $R'=\{(O),(A),(TB),(FB'),(T)\}$ 

\bprop
Consider the rule:
\medskip

{\bf (FB$'$):} Let $a_t\in A$ and $b_h,b_j\in B$ such that  $k_j<t\leq m_j$ and $a_tb_h\in E$. Then $b_h\precf b_j$.
\medskip

Then the sets of rules $R$  and $R'$ generate the same set of forced  pairs $b_h\precf b_j$. 
\label{prop:Fbcaravantmj}
\eprop

\bproof
We show that (FB) is a particular case of (FB'), meaning that  $R'$ generates every forced pair that $R$ generates. Under the hypothesis of the rule (FB), we deduce 
$a_t\precf b_w$, using (A). With $b_w\precf b_j$, we obtain $a_t\precf b_j$, so that $a_t\preceqf a_{m_j}$, which is equivalent with
$t\leq m_j$. Moreover, since $k_j<t$ we have $a_tb_j\not\in E$. Thus (FB) is a particular case of (FB').

Conversely, let $a_t, b_h, b_j$ be chosen as in the hypothesis of (FB'), and let us show that
we can deduce by the rules in $R$ that $b_h\precf b_j$. Since $k_j<t\leq m_j$, we have that $a_tb_j\not\in E$, and also that 
$a_{m_j}b_j\not\in E$. Then Proposition \ref{prop:situationsonA} for  $a_{m_j}\precf b_j$ implies that only case (b) can hold, 
since case c) is impossible by the maximality of $m_j$. So there exists a neighbor $b_u$ of $a_{m_j}$ such that $b_u\precf b_j$. Note that $m_j=k_u$.

Among all the $B$-origins $b_y$ with $b_u\preceqf b_y\precf b_j$ and $t\leq k_y$, let $b_z$ be
closest to $b_j$ with these properties, with respect to $\precf$. Then $b_u\preceqf b_z$. 
If $a_tb_z\in E$, then by (FB) with $b_z$ instead of $b_w$ we deduce that $b_h\precf b_j$ and we are done. Otherwise, 
$a_tb_z\not\in E$ and $t< k_z$. Proposition \ref{prop:3situations} implies that $b_z$ is in one of the cases (1)-(4) 
with respect to $b_j$. 

Case (1) implies the existence of a neighbor $a_r$ of $b_z$ such that $r< k_j$. Then, since $t<k_z$, (TB) may be applied 
to deduce $b_h\precf b_z$ and thus by (T) $b_h\precf b_j$.

Case (2) implies the existence of a basis $b_w$ of $b_j$ such that $b_z$ is a descendant of $b_w$. Then, $a_tb_w\in E$
implies as before by (FB) (with $b_w$ instead of $b_z$) that $b_h\precf b_j$, whereas $a_tb_w\not\in E$ and $t<k_w$
implies, by (TB) and (T) as in case (1), that $b_h\precf b_j$. We study the case where $a_tb_w\not\in E$ and $k_w<t$.
Proposition \ref{prop:status2} then implies, with the same notations, that $t$ belongs to one of the intervals
$I_1=[a_{k_w},a_1]$, $I_r=[a_{u_{r-1}}, a_{u_r}]$ for $2\leq r\leq g-1$ and $I_g=[a_{u_{g-1}},b_{u_g}]$, where $b_{u_g}=b_z$,
which cover all the interval $[a_{k_w},b_z]$. Say that $t\in I_r$ for some $r$. Then either $a_tb_{u_r}\in E$ and we conclude
by (FB) that $b_h\precf b_{u_j}$, or  $a_tb_{u_r}\not\in E$ and $a_t$ is placed between two neighbors of $b_{u_r}$. These neighbors
are the limits of the interval $I_r$ when $r<g$, by Proposition  \ref{prop:status2}, and are $a_{u_{g-1}}$ and $a_{k_z}$
when $r=g$. But then by (TB) we have that $b_h\precf b_{u_r}$ and by (T) that $b_h\precf b_j$.

In cases (3) and (4),  let $b_q$ be the $B$-origin such that $b_q\precf b_j$ and $b_z$ is either a basis or a descendant
of a basis of $b_q$. Thus $b_z\precf b_q\precf b_j$  and we cannot have $t\leq k_q$ by the choice of $b_z$, as
close as possible to $b_j$ with respect to $\precf$ such that $t\leq k_z$. Thus $k_q<t$. But then cases (3) and (4) are respectively
similar to cases (1) and (2) when $b_j$ is replaced with $b_q$. We only have to show that $t\leq m_q$ and this is true
since $m_q=m_j$. Indeed, we cannot have $m_q>m_j$ since then we would have $a_{m_j}\precf a_{m_q}\precf b_q\precf b_j$
which contradicts the definition of $m_j$. And we cannot have $m_q<m_j$ since $a_{m_j}b_u\in E$ and $b_u\precf b_q$.
Thus as in (1) and (2) we deduce that $b_h\precf b_q$ and by (T) we conclude. 
\eproofs

Let $\prec$ be a total order on $A\cup B$ that extends $\precf$, assuming $\precf$ is a partial order.  
A $B$-origin $b_j$ with the property that $a_{m_j}\prec b_j\prec a_{m_j+1}$ is called {\em left-optimal} in $\prec$. Intuitively,
among all the possible places for $b_j$, in the Stick representation $\prec$ its place is chosen as close to the left as possible, 
up to the places of the $B$-origins which are not comparable with it and belong to the same region.
Then we say that $(b_h,b_j)$ is a {\em weak forced pair}, denoted as $b_h\precw b_j$, if $b_h\not\precf b_j$,
$b_j\not\precf b_h$ and $m_h<m_j$, meaning that the order
$b_h\prec a_{m_j}\prec b_j$ is exclusively due to the left-optimal placement of  $b_h$ and $b_j$ in $\prec$. A Stick representation in which all $B$-origins are 
left-optimal is called {\em left-optimal}. Finally, the order $\prec$ is called {\em canonical} if it extends both $\precf$ and 
$\precw$ to $A\cup B$, {\em i.e.} if it extends $\precf$ and is left-optimal.

We are ready now to give and prove the main result.

\bthm
The graph $G=(A\cup B, E)$ has an $A$-Stick representation if and only if the forced pairs $b_h\precf b_j$ obtained from the 
order on $A$ only  using the rules in $R'$ (or, equivalently, in $R$) define a partial order $\precf$. Moreover, in this case any 
canonical order $\prec$ induces an $A$-Stick representation of $G$.  
\label{thm:main}
\ethm

\bproof
The forward direction is obvious. For the reverse direction, we consider a total order $\prec$ extending $\precf$ and such that each $B$-origin 
is left-optimal (that is, $\prec$ is a canonical order). We attempt a Stick representation as follows. Each (horizontal) $A$-segment 
$A_i$ is defined to start in position $a_i$ and to have its tip on the (vertical) 
segment $B_t$, where $b_t=\max\limits_\prec\{b_u\, |\, a_ib_u\in E\}$.  Similarly, 
each (vertical) segment $B_j$ starts in position $b_j$ and has its tip on  the (horizontal) 
segment $A_h$, where $a_h=\min\limits_{\precf}\{a_g\, |\, a_gb_j\in E\}$. In this representation, if $a_ib_j\in E$, then segments 
$A_i$ and $B_j$ necessarily intersect. It remains to show that whenever $a_ib_j\not\in E$, the segments $A_i$ and $B_j$ do not
intersect.

Assume by contradiction that we may have $a_i$ and $b_j$ such that $a_ib_j\not\in E$, but the segments $A_i$ and $B_j$
intersect.  
Then, by the abovementioned construction of the segments, there exists $b_h$ with 
$b_j\prec b_h$ such that $a_ib_h\in E$, and $a_v$ with $a_v\precf a_i$ such that $a_vb_j\in E$. 

Now, $a_{k_j}\precf a_i$, otherwise (TB) with $a_s=a_v$ and $a_t=a_i$ (see Remark \ref{rem:successive}) would imply $b_h\precf b_j$, which is either in contradiction
with the assumption that $\prec$ has no circuit (when $b_j\precf b_h$ also holds) or with the assumption that $\prec$ extends $\precf$ (in the
contrary case). Furthermore, $b_j$ is left-optimal in $\prec$, thus it is placed between $a_{m_j}$ and $a_{{m_j}+1}$. We cannot have
$a_{m_j}\precf a_i$ since then we would have $a_{{m_j}+1}\preceqf a_i$ and $b_j$ should be placed before $a_i$ in $\prec$, by the left-optimality of $b_j$. 
Thus $i\leq m_j$. Putting all these deductions altogether, we have that $k_j<a_i\leq m_j$ and $b_ha_i\in E$. But then by (FB')
we deduce that $b_h\precf b_j$, a contradiction. 
\eproofs

\br
It is important to note here that Theorem \ref{thm:main} only claims that the forced pairs detected by the five rules in $R'$ 
are sufficient to  deduce whether $G$ is an $A$-Stick graph or not. The theorem does not claim that all the forced pairs
are obtained by these rules, nor that {\em each} order extending $\precf$ yields an $A$-Stick representation. Some forced 
pairs $(b_u,b_v)$ are not detected by the rules, thus they do not satisfy $b_u\precf b_v$. However, they are ranged in the 
correct order by the canonical order since they are detected as weak forced pairs,  $b_u\precw b_v$.
\er

\bex
Consider the graph in Figure \ref{fig:exStick} (left) and assume the order $a_1, a_2, a_3, a_4, a_5$ is fixed.
Then by (TB) we deduce that $b_2\precf b_3$ and $b_4\precf b_3$. Moreover, by (T) for $a_4\precf b_2\precf b_3$ it may be deduced that $a_4\precf b_3$
and therefore by (FB') that $b_1\precf  b_3$. No rule allows to deduce $b_4\precf b_2$. However, if
$b_4$ is placed between $b_2$ and $b_3$, the adjacency matrix of $G$ ordered accordingly contains the configuration
$P_3$ in Theorem~\ref{thm:deLuca}, given by $a_2, a_3, b_2, b_4, b_3$, a contradiction. Thus $b_4$ cannot be placed after $b_2$,
implying that $(b_4, b_2)$ is a forced pair, not detected by the forcing rules. In this case, the canonical order will place
correctly $b_4$ before $b_2$ since $m_4=2<4=m_2$, thus $b_4\precw b_2$ is identified as a weak forced pair.
\label{ex:no1}
\eex

The algorithm we give in the next section deduces the forced pairs $b_h\precf b_j$ using $R'$, and has the 
particularity to consider only {\em increasing} forcing sequences. A forcing sequence is {\em increasing} if the
sequence of supports used by its rules is an increasing sequence (not necessarily strictly increasing). 
The correctness of this approach is shown below:

\bthm
Let $b_h\precf b_j$. Then there exists an increasing forcing sequence for it using the rules in $R'$ (similarly for $R$).
\label{thm:increasing}
\ethm

\bproof
Corollary \ref{cor:inductive12} allows us to deduce that there exists a sequence
$q_1, q_2, \ldots, q_p$ of distinct integers such that $q_1=h, q_p=j$ and $b_{q_{r}}$ is either a basis or a descendant
of a basis $b_{q_{r+1}}$, for $1\leq r\leq p-1$. Let $S_r$ be the forcing sequence attesting that $b_{q_r}$ is a basis or a descendant
of a basis of $b_{q_{r+1}}$ (we may assume we use (FB') whenever we should use (FB) to obtain a descendant). Then, for each $r\neq r'$:

\begin{itemize}
 \item  $S_r$ is increasing, trivially when  $b_{q_{r}}$ is a basis  of $b_{q_{r+1}}$ since the length of $S_r$ is 1, and
by Proposition \ref{prop:status2} when $b_{q_{r}}$ is  a descendant of $b_{q_{r+1}}$.
 \item $S_r$ and $S_{r'}$ are disjoint, since $S_r$ contains only rules deducing $b_x\precf b_{q_{r+1}}$ (for 
 appropriate $b_x$), whereas $S_{r'}$ contains only rules deducing $b_y\precf b_{q_{r'+1}}$ (for appropriate $b_y$), and 
 $b_{q_{r+1}}\neq b_{q_{r'+1}}$. 
 \item $S_r$ and $S_{r'}$ have independent rules, in the sense that no rule in $S_r$ uses a result obtained by a rule in $S_{r'}$,
 nor viceversa. The argument is similar to the previous one.
\end{itemize}

Then we can build an increasing forcing sequence $S$ for $b_h\precf b_j$ using the rules in $R'$
by merging $S_1, \ldots, S_{p-1}$.
\eproof

\section{A linear algorithm for STICK$_A$}\label{sec:linalg}

We now propose an $O(|A|+|B|+|E|)$ running time algorithm that, given a bipartite graph $G=(A\cup B)$ and an order $\precf$ on $A$, computes the 
forced pairs implied by the forcing rules in $R'$, tests whether they form a partial order and, simultaneously, computes the
weak forced pairs. The result is either a pair of vertices that are forced in both directions (implying a circuit
in the order $\precf$) or a partial order all of whose extensions are canonical orders.

To this end, we need some definitions. A {\em bubble} is a connected directed acyclic graph with a unique source $s$,
a unique target $t\neq s$ and such that each vertex  is  on a directed path from $s$ to $t$.
We say that the source $s$ of a bubble {\em opens} the bubble, whereas its target {\em closes}
the bubble. The bubble is called {\em trivial} when it contains no vertex but $s$ and $t$, which are 
then joined by an arc from $s$ to $t$.  It is called {\em linear} when it is a sequence of linearly ordered induced (sub-)bubbles,
the target of each bubble being the source of the next bubble (except for the last bubble).

\subsection{The algorithm}\label{subsec:thealg}
The algorithm computes a directed acyclic graph (DAG) $D$ such that $x$ precedes $y$, with $x,y\in A\cup B$, if and only if 
$x\precf y$ or $x\precw y$.
At the end of each execution $i$ of the main loop, the following invariants are verified:

\begin{itemize}
\item[(I.1)] the vertices of $D$ are given by $V(D)=B_i\cup J_i$, where $B_i=\bigcup_{t=1}^i N(a_t)$ and $J_i$ is made of 
supplementary vertices called {\em connectors}, $i$ of which are labeled $a_1$, \ldots, $a_i$. Some connectors are
{\em strong}. Their role is to open and close the bubbles. 
\item[(I.2)] $D$ is a linear bubble, made of at most $i+1$ so-called {\em main} bubbles $X_0, X_{t_1}, \ldots, X_{t_v}$,  
$1\leq t_1<\ldots <t_v\leq i$, with the following properties:
\begin{itemize}[leftmargin=1cm]
 \item[(I.2a)] their linear order from left to right (or from the source of $D$ towards its target) is $X_0, X_{t_v}, X_{t_v-1},$ $\ldots, X_{t_1}$. 
 \item[(I.2b)] the source of the bubble $X_{t_k}$ is the so-called {\em strong} connector $s_{t_k}$, whereas its 
 target is the source $s_{t_{k+1}}$ of the next bubble $X_{t_{k+1}}$ according to the linear order, except for $X_{t_1}$ whose target is  
 the final strong connector $s_{0}$. The strong connector $s_{t_v}$, which separates the {\em settled zone} 
 $X_0$ and the {\em working zone} $D-X_0$ is also denoted $First$.
\item[(I.2c)] $X_0$ is a linear bubble with source $a_1$ (a strong connector) and target $First$. It contains possibly trivial {\em small} bubbles 
$Y_1, \ldots, Y_p, \ldots Y_{i}$ in increasing order of $p$ from left to right, where $p$ spans all the integers $1, 2, \ldots, i$ but also some intermediate, 
non-integer, values. The source of $Y_j$ is the strong connector $a_j$ and its target is the next strong connector $a_{j'}$ 
to the right, where $j'\in\{j'+\frac{1}{2},j'+1\}$, except for $Y_i$ whose target is $First$.  The $B$-origins in $X_0$ are considered as {\em settled}. 
\item[(I.2d)] $X_{t_k}$, $t_k\geq 1$ is a bubble containing the $B$-origins $b_j$ that are adjacent to $a_{t_k}$, 
to some $a_{x}$ with $x>t_k$ and (if $t_k\neq i$) to no
$a_y$ such that $t_k<y\leq i$.  
\end{itemize}

\item[(I.3)] each path in $D-X_0$, as well as each path in the small bubbles, alternates $B$-origins and connectors.
\item[(I.4)] the indegree and outdegree of each $B$-origin $b_j$ in $D$ is 1, with arcs respectively from the {\em previous connector} 
of $b_j$ denoted $\Prev(b_j)$ and to the {\em next connector} of $b_j$ denoted $\Next(b_j)$.
\end{itemize}
\medskip

For each $i$, the directed paths in $D$ are aimed at 1) recording all the forced pairs deduced from forcing rules 
with supports $a_f\preceqf a_i$, and 2) indicating the left-optimal places for the elements $b_j$ for which $m_j\leq i$.

In our algorithm, rules (TB) and (FB') are treated similarly. Note that, when it implies that $b_h\precf b_j$, the rule (TB) exploits 
the existence of a neighbor $a_t$ of $b_h$ between two successive neighbors of $b_j$ (see Remark~\ref{rem:successive}), whereas 
(FB') exploits the existence of a  neighbor $a_t$ of $b_h$ between the neighbor $a_{k_j}$
of $b_j$ (excluded) and $a_{m_j}$ included (assuming $a_t\precf a_{m_j}$ is already known). As a consequence, the two rules are similar if we accept,
when $k_j<m_j$, to  artificially add a new, artificial, neighbor of $b_j$  between $a_{m_j}$ 
(in fact, between its approximation at any given time) and $b_j$.
Then, each time the largest $y$ such that $a_y\precf b_j$
is updated by the algorithm, an artificial $A$-origin named $a_{y+\frac{1}{2}}$ is created and it becomes the largest (artificial) neighbor of $b_j$,
thus replacing $a_{m_j}$ when the rule (FB') is applied. In consequence, (TB) and (FB') can be treated similarly. 
Note that only the largest artificial neighbor of $B_j$ is necessary for (FB'), so this is the only one we have to record. Moreover, 
when $a_y\precf b_j$ is detected, this also implies that $a_y\precf b_{j'}$ for all $b_{j'}$ such that $b_j\precf b_{j'}$, and therefore the new neighbor 
$a_{y+\frac{1}{2}}$ must also be forwarded to all these $B$-origins $b_{j'}$. 

The artificial $A$-origins $a_{y+\frac{1}{2}}$ are treated similarly to the initial $A$-origins $a_i$, and have their dedicated steps.
\medskip

 {\bf Intuitive description of the algorithm CSO.} 
The algorithm scans the $A$-origins $a_i$, with $i$ integer or not, and performs a step $i$ whose aims are: 1) to identify the 
forced pairs $(b_h,b_j)$ for which (TB) or (FB') can be applied with support $a_i$; 2) to place $b_h$ as a predecessor of 
$b_j$ in $D$ (the rule (T) is then implicitly applied for the successors of $b_j$); and 3) for each weak forced pair $(b_h,b_j)$ 
for which $m_h=\lfloor i\rfloor$, to place $b_h$ after $a_{m_h}$ and before $a_{m_h+1}$. To this end, step $i$ starts with the DAG $D$
computed at the end of step $i-1$, and performs the following treatment depending on $i$. 

\begin{figure}[t!]
\centering
\vspace*{-2cm}
 \hspace*{-0.6cm}\includegraphics[width=0.95\textwidth]{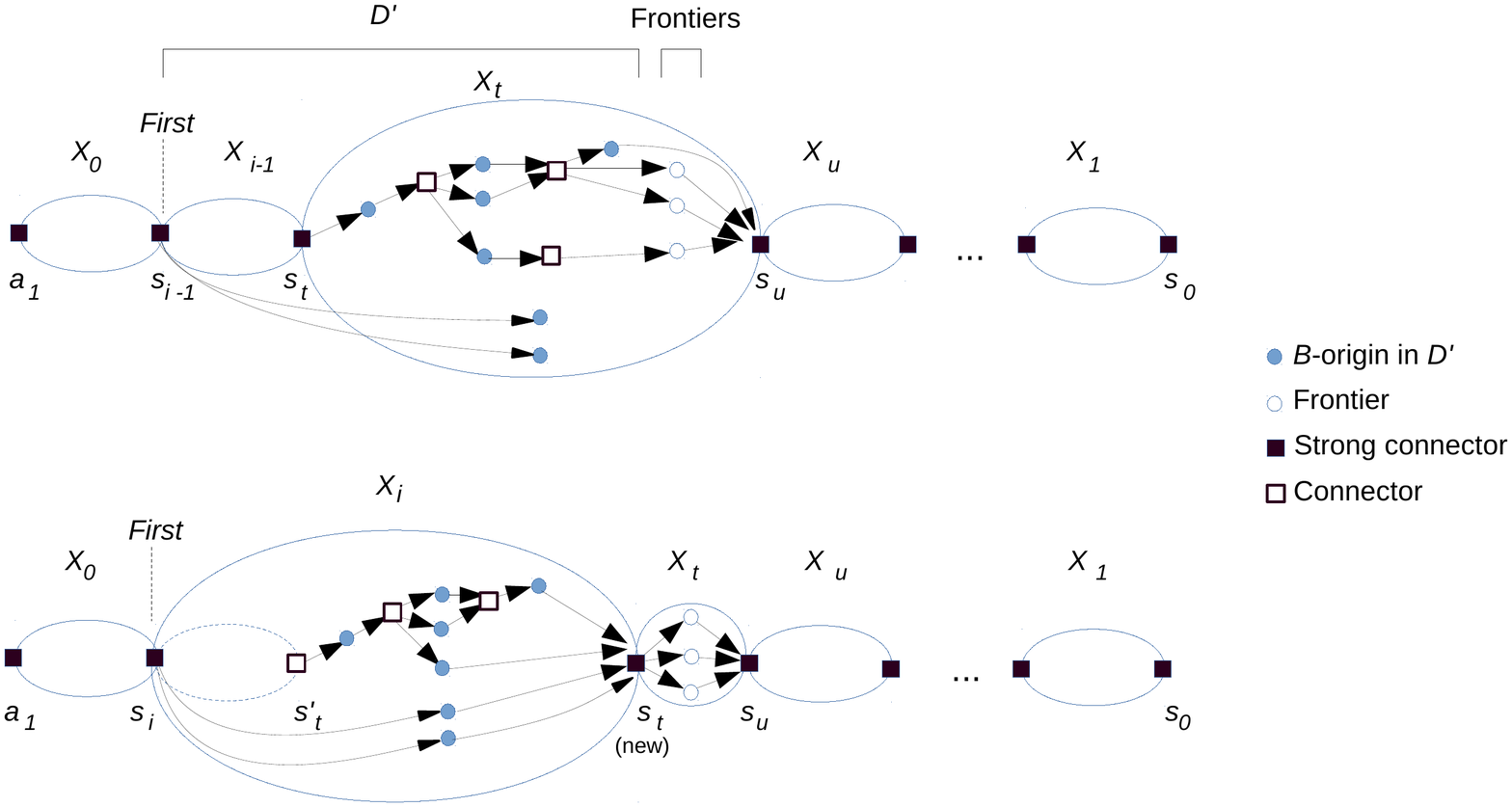}
 \vspace*{-1.5cm}
\caption{\small An example of bubble closing, used in steps 3.$i$.4 and 3.$i$.7 of Algorithm CSO. Top: The graph $D$ at the end 
of step 3.$i$.3. Down: The graph $D$ at the end of step 3.$i$.4. The frontiers
 and their successors (if any) from $X_t$ are kept in $X_t$, whereas the new bubble $X_i$ containing $D'$ is closed using a new connector
 $s_t$. The old strong connector $s_t$ is renamed as $s'_t$, and $X_{i-1}$ is no longer a main bubble.}
 \label{fig:bubbleclosing}
\end{figure}

If $i$ is integer,  the algorithm  identifies in $D$ the minimum induced sDAG $D'$ containing all the vertices in $N(a_i)\cap V(D)$. Its
properties allow us to decide whether a certificate indicating that $G$ cannot be $A$-Stick is found or not (step 3.$i$.2); in the former 
case the algorithms stops. In the latter case, each $B$-origin in $N(a_i)\setminus V(D)$ is added to $D'$ (step 3.$i$.3).
and the algorithm creates (step 3.$i$.4, see also Figure~\ref{fig:bubbleclosing}) a new bubble, containing 
$D'$ only, immediately after the settled zone and before the leftmost $B$-origins in $D-X_0-D'$ (called {\em frontiers}). The
result is that all the vertices $b_h$ in $N(a_i)$ precede in the new DAG $D$ all the vertices $b_j$ in $D-X_0-D'$,
that is, all the vertices $b_j$ that are non-adjacent to $a_i$ but have neighbors (true or artificial) before $a_i$ and 
after $a_i$. This placement of $D'$ before $D-X_0-D'$ realizes the applications of rules   
(TB) and (FB') to $b_h$ from $D'$ and $b_j$ from $D-X_0-D'$, with support $a_i$.  

For each vertex $b_j$ in $D-X_0$, its neighbor with largest index (integer or not) is then computed or updated (step 3.$i$.5). 
This index is the current approximation of $m_j$, and is stored in the variable $\Last_j$. More precisely, 
$\Last_j$ is initially set to $k_j$ and this is the only integer value of $\Last_j$. When $\Last_j$ is not an integer, $\Last_j=y+\frac{1}{2}$ where 
$y>k_j$ and $a_y$ is the maximum neighbor, with respect to $\precf$,  of a predecessor $b_h$ of $b_j$ in $D-X_0$ ({\em i.e.} $a_y=a_{k_h}$).
In contrast with the description of step 3.$i$.5 in the algorithm CSO, if we want the algorithm to have a linear running time, 
$\Last_j$ cannot be updated for  each $b_j$ at each step of the algorithm (the necessary modifications are shown later).

Further, whether $i$ is integer or not, the $B$-origins $b_h$ for which $a_i$ is the neighbor with largest 
index, meaning that $\Last_j=i$ and $\lfloor i\rfloor=m_h$,  are identified (step 3.$i$.6). The left-optimality requires
these $B$-origins to be placed after $a_{\lfloor i\rfloor}$ and before $a_{\lfloor i\rfloor+1}$, so we
place them in the bubble $Y_i$  created at the end of the settled zone, and before the working zone  
(step 3.$i$.7, similar to 3.$i$.4).  
\bigskip

In Algorithm CSO (framed), a {\em compact sDAG} $H$ of $D$ is an induced subgraph of $D$ with the property that each vertex
on a path between two vertices in $H$ also belongs to $H$. The variable $First$ is seen as a pointer, which - when
the name of the connector it points to changes - continues to point to the same connector.
\newpage

\begin{framed}

{\small
\noindent{\bf Algorithm CSO (Canonical $A$-Stick Order)}

\noindent{\bf Input:} A bipartite graph $G=(A\cup B, E)$ without isolated vertices in $B$, an order $\precf$ on $A$.

\noindent{\bf Output:} A canonical $A$-representation of $G$ if $G$ is $A$-Stick; the answer ``$G$ is not $A$-Stick'' otherwise.
\medskip

\begin{enumerate}[itemsep=0.1cm,topsep=0cm,parsep=0cm]
 \item[1.] Let $D=(\{s_0\},\emptyset)$ be a DAG with a unique strong connector. Let $First=s_0$ and $X_0=D$.
 \item[2.] For each $b_j\in B$, let $\Last_j=k_j$. 
\item[3.] Perform the following operations for each $a_i$ in increasing order of $i$, including the artificial $A$-origins with a non-integer $i$,
that are progressively inserted by the algorithm.

If $i$ is integer then:

\begin{enumerate}[itemsep=0.0cm,topsep=0cm,parsep=0cm,leftmargin=1.2cm]
 \item[3.$i$.1] Let $N(a_i)=N^D(a_i)\cup N^{\overline{D}}(a_i)$ be the partition of $N(a_i)$ into $B$-origins that already belong to $D$ 
 (in fact, to $D-X_0$) and $B$-origins that do not belong to $D$
 \item[3.$i$.2] Let $D'$ be the induced sDAG of $D$ with vertices $\{First\}\cup N^D(a_i)\cup \{\Prev(b_j)\, |\, b_j\in N^D(a_i)\}$.
 If $D'$ is not compact, return ``$G$ is not $A$-Stick''. 
 \item[3.$i$.3]  Add to $D'$, and thus to $D$, the vertex set $N^{\overline{D}}(a_i)$ and arcs from $First$ to each vertex in $N^{\overline{D}}(a_i)$. 
\end{enumerate}

\hspace*{0.3cm} If  $V(D')\neq \{First\}$:

\begin{enumerate}[itemsep=0.0cm,topsep=0cm,parsep=0cm,leftmargin=1.25cm]

 \item[3.$i$.4]  Transform $D$ by closing the bubble $X_i$ of $D'$ as follows. (See Figure~\ref{fig:bubbleclosing}.) Let $s_t$ be the rightmost strong connector of $D'$,
 and $s_u$ be the next strong connector to the right (if any).
 If there is some $B$-origin in $V(X_t)\setminus V(D')$, then remove from $D$ all the next connectors 
 of the targets in $D'$ except $s_u$, rename $s_t$ as $s'_t$ and create a new strong connector $s_t$. Then let $s_t$ be the unique 
 out-neighbor of all the targets in  $D'$ and the unique in-neighbor of all the $B$-origins in $V(X_t)\setminus V(D')$ whose previous 
 connector belongs to $D'$ or has been removed. On the contrary, if  there is no $B$-origin in $V(X_t)\setminus V(D')$ and $i\neq 1$, just add an arc from each 
 vertex  in $N^{\overline{D}}(a_i)$ towards $s_u$; when $i=1$, rename $s_0$ as $s_1$, create a new strong connector $s_0$ and add
 an arc from each vertex in $N(a_i)$ towards $s_0$.
 In all cases, let $First$ be renamed as $s_i$.  The former strong connectors of $D'$, except $First$, 
 become simple connectors. 
 
\item[3.$i$.5] Update $\Last_j$ for all $b_j$ to the right of $First$ as follows:
\medskip

\hspace*{1cm} $y_{b_j}=\max\{\Last_x\, |\, First\, \hbox{precedes}\, b_x\, \hbox{which precedes}\, b_j\, \hbox{in}\, D\}.$
\medskip

If 
$\Last_j<y_{b_j}$ then
either $y_{b_j}$ is an integer, in which case let $\Last_j=y_{b_j}+\frac{1}{2}$ (create $a_{y_{b_j}+\frac{1}{2}}$ if it does 
not exist yet); 
or $y_{b_j}$ is not an integer, and then let $\Last_j= y_{b_j}$. In both cases, let $b_j$ be adjacent to $a_{\Last_j}$
and remove the previous adjacency (if any) between $b_j$ and an artificial $A$-origin.

\end{enumerate}

\hspace*{0.3cm} Endif.

Endif.

\begin{enumerate}[itemsep=0.0cm,topsep=0cm,parsep=0cm]

 \item[3.$i$.6] Let $F'$ be the compact induced sDAG of $D$ with vertices $\{First\}\cup\{b_j\, |\, \Last_j=i\}\cup \{\Prev(b_j)\, |\, \Last_j=i\}$.

\item[3.$i$.7]   If $V(F')\neq\{First\}$, then transform $F'$ into a bubble of $D$ by closing it at follows. 
Let $s_q$ be the rightmost strong connector of $F'$ and $s_w$ be the next strong connector to the right.
 If there is some $B$-origin in $V(X_q)\setminus V(F')$, then remove from $D$ all the next connectors 
 of the targets in $F'$ except $s_w$, rename $s_q$ as $s'_q$ and create a new strong connector $s_q$. Then let $s_q$ be the unique 
 out-neighbor of all the targets in  $F'$ and the unique in-neighbor of all the $B$-origins in $V(X_q)\setminus V(F')$ whose previous 
 connector belonged to $F'$ or has been removed. On the contrary, if there is no $B$-origin in $V(X_q)\setminus V(F')$, then the bubble of $F'$ is already closed. 
 The former strong connectors of $F'$, except $First$, become simple connectors.
 Let the vertex $First$ be renamed with $a_i$. Update $First$ to be the strong connector closing the bubble of $F'$.
 
 If $V(F')=\{First\}$, then insert a strong connector $a_i$ between $First$ and its in-neighbors, so that $a_i$ becomes the unique in-neighbor
 of $First$. 
 
 Let $Y_i$ be the trivial bubble opened by $a_i$ and closed by $First$.

 \end{enumerate}

 \item[4.] 
 Let $\preccso$ be the partial order on $A\cup B$ defined as
$c_x\preccso c_y$ if and only if $c_x$ precedes $c_y$ in $D$.
Return any order $\prec$ which extends the partial order $\preccso$.
  
\end{enumerate}
}

\end{framed}

We propose an example demonstrating the algorithm.

\bex  We consider again the graph in Figure \ref{fig:exStick} with the order $a_1, a_2, a_3, a_4, a_5$
In the execution of our algorithm (see Figure \ref{fig:example}), we mark the position of $First$ by a $*$,
and write in bold the strong connectors. The numbers in parentheses represent the values $\Last_j$, for the $B$-origins $b_j$.
As indicated in Example \ref{ex:no1}, we have $b_1\precf  b_3$, $b_2\precf b_3$ and $b_4\precf b_3$, which
induce a partial order, thus the graph is $A$-Stick. The left-optimality
ranges $b_4, b_2$ and $b_1$ in this order from left to right, immediately after the $A$-origins defining their
$m_h$ values ($h=4, 2, 1$). The order output by the algorithm 
on the last line in Figure \ref{fig:example} (just forget $a_{5\frac{1}{2}}$ and $s_0$) is thus the expected one.

\begin{figure}[t]
\centering
    \resizebox{0.97\textwidth}{!}{
\begin{tabular}{cclcl}
 $i$&Step &DAG& $s_t/s_q$&Particular case in bubble closing\\ \hline
 1&3.1.3 &$D':{\bf s_0^*}\rightarrow b_3$&${\bf s_0}$&\\
 &&\hspace*{0.95cm} \searrownew\, $b_4$&&\\
 &3.1.5&$D: {\bf s_1^*}\rightarrow b_3(3)\rightarrow {\bf s_0}$&&$V(X_{s_0})\setminus V(D')=\emptyset, i=1$\\
  &&\hspace*{0.85cm} \searrownew\,  $b_4(2)$ $\nearrownew$&&\\
 &3.1.6&$F':{\bf s_1^*}$& ${\bf s_1}$&\\
 &3.1.7&$D: {\bf a_1}\rightarrow {\bf s_1^*}\rightarrow b_3(3)\rightarrow {\bf s_0}$&&$V(F')=\{First\}$\\ 
 &&\hspace*{1.8cm} \searrownew\,  $b_4(2)$ $\nearrownew$&&\\\hline
 2&3.2.3&$D':{\bf s_1^*}\rightarrow b_4(2)$&${\bf s_1}$&\\
  &&\hspace*{0.95cm} \searrownew\,  $b_2$&&\\
 &3.2.5&$D: {\bf a_1}\rightarrow {\bf s_2^*}\rightarrow b_4(2)\rightarrow {\bf s_1}\rightarrow b_3(3)\rightarrow {\bf s_0}$&&$V(X_{s_1})\setminus V(D')\supseteq\{b_3\}\neq \emptyset$\\
  &&\hspace*{1.75cm} \searrownew\,  $b_2(4) \nearrownew$&&\\
  &3.2.6& $F':{\bf s_2^*}\rightarrow b_4(2)$ &${\bf s_2}$&\\
  &3.2.7&$D: {\bf a_1}\rightarrow {\bf a_2}\rightarrow b_4(2)\rightarrow {\bf s_2^*}\rightarrow b_2(4)\rightarrow {\bf s_1}\rightarrow b_3(3)\rightarrow {\bf s_0}$&&$(V(X_{s_2})\setminus V(F'))\cap B=\emptyset$\\ \hline
3&3.3.3&$D':{\bf s_2^*}\rightarrow b_2(4)\rightarrow {\bf s_1}\rightarrow b_3(3)$&${\bf s_1}$&\\
&3.3.5&$D: {\bf a_1}\rightarrow  {\bf a_2}\rightarrow b_4(2)\rightarrow {\bf s_3^*}\rightarrow b_2(4)\rightarrow s'_1\rightarrow b_3(4\frac{1}{2})\rightarrow {\bf s_0}$&&$V(X_{s_1})\setminus V(D')=\emptyset, i\neq 1$\\
&  3.3.6&$F':{\bf s_3^*}$&${\bf s_3}$&\\
&  3.3.7&$D: {\bf a_1}\rightarrow {\bf a_2}\rightarrow b_4(2)\rightarrow {\bf a_3}\rightarrow {\bf s_3^*}\rightarrow b_2(4)\rightarrow s'_1\rightarrow b_3(4\frac{1}{2})\rightarrow {\bf s_0}$&&$V(F')=\{First\}$\\ \hline
4&3.4.3&$D':{\bf s_3^*}\rightarrow b_2(4)$&${\bf s_3}$&\\
&&\hspace*{0.95cm} \searrownew\,  $b_1(5)$&&\\
&3.4.5&$D: {\bf a_1}\rightarrow {\bf a_2}\rightarrow b_4(2)\rightarrow {\bf a_3}\rightarrow {\bf s_4^*}\rightarrow b_2(4)\rightarrow {\bf s_3}\rightarrow b_3(5\frac{1}{2})\rightarrow {\bf s_0}$&&$V(X_{s_3})\setminus V(D')\supseteq\{b_3\}\neq\emptyset$\\
&&\hspace*{5.15cm} \searrownew\,  $b_1(5)\nearrownew$&&\\
&3.4.6&$F':{\bf s_4^*}\rightarrow b_2(4)$&${\bf s_4}$&\\
&3.4.7&$D: {\bf a_1}\rightarrow {\bf a_2}\rightarrow b_4(2)\rightarrow {\bf a_3}\rightarrow {\bf a_4}\rightarrow b_2(4)\rightarrow {\bf s_4^*}\rightarrow b_1(5)\rightarrow {\bf s_3}\rightarrow b_3(5\frac{1}{2})\rightarrow {\bf s_0}$&&$V(X_{s_4})\setminus V(F')\supseteq\{b_1\}\neq\emptyset$\\ \hline
5&3.5.3& $D': {\bf s_4^*}\rightarrow b_1(5)$&${\bf s_4}$&\\
&3.5.5&$D: {\bf a_1}\rightarrow {\bf a_2}\rightarrow b_4(2)\rightarrow {\bf a_3}\rightarrow {\bf a_4}\rightarrow b_2(4)\rightarrow {\bf s_5^*}\rightarrow b_1(5)\rightarrow {\bf s_3}\rightarrow b_3(5\frac{1}{2})\rightarrow {\bf s_0}$&&$(V(X_{s_4})\setminus V(D'))\cap B=\emptyset$\\
&3.5.6&$F': {\bf s_5^*}\rightarrow b_1(5)$&${\bf s_5}$&\\
&3.5.7&$D: {\bf a_1}\rightarrow {\bf a_2}\rightarrow b_4(2)\rightarrow {\bf a_3}\rightarrow {\bf a_4}\rightarrow b_2(4)\rightarrow {\bf a_5}\rightarrow b_1(5)\rightarrow {\bf s_3^*}\rightarrow b_3(5\frac{1}{2})\rightarrow {\bf s_0}$&&$(V(X_{s_4})\setminus V(F'))\cap B=\emptyset$\\ \hline
$5\frac{1}{2}$&$3.5\frac{1}{2}$.6&$F': {\bf s_3^*}\rightarrow b_3(5\frac{1}{2})$&${\bf s_3}$&\\
&$3.5\frac{1}{2}$.7&$D: {\bf a_1}\rightarrow {\bf a_2}\rightarrow b_4(2)\rightarrow {\bf a_3}\rightarrow {\bf a_4}\rightarrow b_2(4)\rightarrow {\bf a_5}\rightarrow b_1(5)\rightarrow {\bf a_{5\frac{1}{2}}}\rightarrow b_3(5\frac{1}{2})\rightarrow {\bf s_0}$&&$(V(X_{s_3})\setminus V(F'))\cap B=\emptyset$\\ \hline
  \end{tabular}
  }
  \caption{An example of execution.}
  \label{fig:example}
  \end{figure}
\eex

The proof of correctness for Algorithm CSO follows. The running time of the algorithm is the object of  Subsection \ref{subsec:runningtime}. 

\subsection{Correctness}\label{subsec:correctness}

Due to Theorem \ref{thm:increasing}, the correction of the algorithm is quite intuitive. However, the formal proof is not immediate.

\br At the end of step 3.$i$ of Algorithm CSO, the digraph $D$ satisfies the invariants (I.1) to (I.4) in the 
description of the graph $D$. They are immediate consequences of the operations performed by the algorithm, except for 
invariant (I.2d). To show it, note that if $b_j$ belongs to the bubble $X_{s_f}$ at the
end of step $i$, that means $b_j$ is adjacent to $a_f$ (by the construction in steps 3.$f$.2 to 3.$f$.4) and has not been
extracted from $X_{s_f}$ then inserted into some other bubble, thus it is not adjacent to another $A$-origin between $a_f$ and $a_i$ ($a_i$ included). However, 
$\Last_j>i$ otherwise $b_j$ would have been moved to $X_0$ in step 3.$\Last_j$.7, thus - by the updates in step 3.$i$.5 - $b_j$ has a neighbor, which 
may be an  artificial one, larger than $a_i$.
\label{rem:invariants}
\er

In each step 3.$i$, we denote $D''$ the induced sDAG $D-D'-X_0$ obtained at the end of step 3.$i$.4 and $F''$ the induced sDAG $D-F'-X_0$ obtained at the 
end of step 3.$i$.7. An important remark is the following:

\br
In Algorithm CSO, if $b_h$ precedes $b_j$ at the end of some step $i$, 
then $b_h$ precedes $b_j$ until the end of the algorithm, whether the algorithm stops in step 3.$u$.2, for some $u$, or in step 4. 
\label{rem:alwaysprecede}
\er

\br
In Algorithm CSO,  for each $b_q$ the value $\Last_q$ is initially integer and equal to $k_q$. If
$\Last_q=t+\frac{1}{2}$,  where $t$ is integer and $t\geq i$, at the beginning of some step 3.$i$ then $\Last_q$ has been necessarily 
updated in some step 3.$g$.5 with $g<i$, which implies that $b_q$ had in $D-X_0$ of step 3.$g$.5 a predecessor $b_r$ with 
$\Last_r\in \{t,t+\frac{1}{2}\}$. A recursive reasoning in case $\Last_r = t+\frac{1}{2}$, combined with  
Remark \ref{rem:alwaysprecede}, imply  the existence of a path $b_{r_1}, b_{r_2}, \ldots, b_{r_p}=b_q $ 
in $D-X_0$ at the beginning of step 3.$i$, with $\Last_{r_1}=t$ and $\Last_{r_i}=t+\frac{1}{2}$ for $2\leq i \leq p$. 
Recall that $i\leq t$, so that no $b_{r_i}$ is moved to $X_0$ before the step 3.$i$. Moreover, none of the values 
$\Last_{r_i}$, $1\leq i\leq p-1$, changed in the meantime, otherwise $\Last_q$ would have changed too.
\label{rem:predecessort}
\er

\bthm
Algorithm CSO correctly tests whether $G$ is an $A$-Stick graph or not. Moreover, if $G$ is $A$-Stick, then the order returned by 
the algorithm is a canonical order.
\label{thm:correction}
\ethm

We need an important preliminary result, presented in Lemma \ref{lem:P}.
Let $m$ be the largest value for which step 3.$m$ is completely performed by Algorithm CSO before it stops. 
Then $m$ is an integer iff the step $m+\frac{1}{2}$ does not exists, since there is no possible end inside a non-integer step. 
We first show the following property: 

\blem
Let $b_x$ be a $B$-origin such that $\Last_x=r+\frac{1}{2}$ at the end of step $3.u$,  with $u\leq m$ and $r$ integer. 
Assume $u\leq r$.  Then $b_x\in V(D)\setminus V(X_0)$ at the end of step $u$, and either $a_rb_x\not\in E$ or $m< r$. 
\label{lem:3}
\elem

\bproof
Since $\Last_x$, initially equal to $k_x$, has been updated, the vertex $b_x$ is necessarily in $D$. Moreover, 
$b_x$ cannot belong to $X_0$, since then the latest value of $\Last_x$ would be equal to $f$, $f\leq u$, where 3.$f$.7 is the step
that put $b_x$ in $X_0$. But $f\leq u<r+\frac{1}{2}=\Last_x$, a contradiction. Thus $b_x\in V(D)\setminus V(X_0)$ at the
end of step 3.$u$.

Further, we show that $r>k_x$ or $r>m$. Since the first value of $\Last_x$ is $k_x$ and the values of
$\Last_x$ strictly increase by definition, we must have $r\geq k_x$. If $r>k_x$, then $a_rb_x\not\in E$ and the
proof is finished. Otherwise, we necessarily have  $r=k_x$. Following Remark \ref{rem:predecessort}, $\Last_x$ has been updated 
in some step 3.$f$.5, with $f\leq u$, following a path with vertices $b_{z_1}, b_{z_2}, \ldots,  b_{z_p}=b_x$ in this order 
from left to right in $D-X_0$, such that $\Last_{z_1}=r$ and  $\Last_{z_p}=r+\frac{1}{2}$ 
at the end of step 3.$f$.5. We cannot have
$p=2$, since in this case before the update in step 3.$f$.5 we had $\Last_x=k_x=r$ and $y_{b_x}=r$. But no update is performed 
under these conditions. Then $p>2$ and $b_{z_2}$ has $\Last_{z_2}<r$ before step 3.$f$.5, implying that $k_{z_2}<r$. But then in step 3.$r$.2
the algorithm finds that $D'$ is not compact, since $b_{z_2}$ belongs to $D''$ and precedes $b_x$, which belongs to $D'$.
Then the algorithm stops in step 3.$r$, thus $r>m$.
\eproofs

In order to show that the algorithm CSO performs correctly, we need to identify and prove the relationships between vertices preceding
each other in $D$, on the one hand, and forced or weak forced pairs, on the other hand. We consider this issue in the intermediate 
result below, that we later use to conclude. 
\bigskip

\blem 
Let $m$ be the largest index for which step 3.$m$ is completely performed before Algorithm CSO stops.
Then:

\begin{enumerate}[itemsep=0cm,parsep=0cm,topsep=0cm]

\item[$\mathcal{(P.\rm{1})}$] Let $D$ be the DAG obtained at the end of step 3.$i$,
for a fixed $i\leq m$. Then $b_h$ precedes $b_j$ in $D$ as a consequence of steps 3.$g$.4, with $g\leq i$, exclusively   
if and only if $b_h\precf b_j$ and there is an increasing forcing 
sequence for it using $R'$, whose supports $a_f$ in the (TB) and (FB') rules satisfy  $f\leq i$. 

\item[$\mathcal{(P.\rm{2})}$] Let $b_h$ be a $B$-origin. Then: 

\begin{enumerate}
\item[a)] For each step 3.$i$ such that $b_h\in V(D-X_0)$ at the end of step
3.$i$.4, we have  
$\lfloor \Last_h\rfloor \leq m_h$ at the end of step 3.$i$.5.

\item[b)] Moreover, if $\Last_h=i$ in step 3.$i$.7 then 
$\lfloor i\rfloor=m_h$.

\item[c)] Conversely, there exists a unique $i$ with $\lfloor i\rfloor=m_h$, such that $\Last_h=i$ 
in step 3.$i$.7.
\end{enumerate}

\item[$\mathcal{(P.\rm{3})}$] Assume $b_h\not\precf b_j$ and $b_j\not\precf b_h$. Then
$b_h$ precedes $b_j$ in the DAG $D$ obtained at the end of step 3.$m$   if and only if 
 $b_h\precw b_j$ with $m_h\leq m$,  and $b_j$ has at least one neighbor 
$a_s$ with $s\leq m$ and $s$ integer.  In this case, $b_h$ has been placed in $X_0$ in step 3.$i$.7, 
with $\lfloor i\rfloor=m_h$.

\item[$\mathcal{(P.\rm{4})}$] The vertex $a_q$, where $q$ is an integer, precedes $b_j$ in the DAG $D$ obtained at the 
end of step 3.$m$  if and only if  $q\leq m$, $b_j$ has a neighbor $a_s$ with $s\leq m$  and  
$a_q\precf b_j$.
\end{enumerate}
\label{lem:P}
\elem

\noindent {\it Proof of $\mathcal{(P.\rm{1})}$$\Rightarrow$}:

 By hypothesis, a set of arcs built in steps 3.$g$.4, with $g\leq i$, decided that $b_h\precf b_j$. Let $g_{hj}$ 
be the largest index $g$ such that step 3.$g$.4 contributed to deduce that $b_h$ precedes $b_j$. We use an induction
on $g_{hj}$ and notice that for $g_{hj}=1$ the affirmation is trivially true. We then prove it in the general case $g_{hj}>1$, 
assuming it holds for smaller values $g_{h'j'}$. Note that all the indices $g_i$ used here are integers.

Then two situations are
possible for $b_h$ and $b_j$: either they both belong to $D-X_0$ at the end of step 3.$g_{hj}$.4; 
or $b_h$ has been moved into $X_0$ in a previous step 3.$g$.7 such that $b_j\not\in V(D)$, but there exists a sequence of 
$B$-origins $b_{q_1}, b_{q_2}, \ldots, b_{q_r}$ such that $b_{q_1}=b_h, b_{q_r}=b_j$ and $b_{q_p}$ precedes $b_{q_{p+1}}$
in the sDAG $D-X_0$ at the end of some step 3.$g_p$.4, for $1\leq p<r$, where $g_1<g_2<\ldots < g_{r-1}=g_{hj}$. 
Each $b_{q_p}$, $p\leq r-2$, on its turn may have been moved to $X_0$ before the insertion of $b_j$ in $D-X_0$. 
In this sequence, affirmation $\mathcal{(P.\rm{1})}$$\Rightarrow$ is true by inductive hypothesis for all
pairs $(b_{q_p}, b_{q_{p+1}})$ but the last one. Then it is sufficient to prove  affirmation 
$\mathcal{(P.\rm{1})}$$\Rightarrow$ for the last pair 
and the second case follows. Just use transitivity to deduce the relation $b_h\precf b_j$, and 
sort the rules in all the resulting increasing forcing sequences to obtain the sought forcing sequence. 
Now, proving $\mathcal{(P.\rm{1})}$$\Rightarrow$ for the last pair is the same as proving $\mathcal{(P.\rm{1})}$$\Rightarrow$ for the
first situation.

We then show that $\mathcal{(P.\rm{1})}$$\Rightarrow$ holds when $b_h, b_j$ belong to $D-X_0$ in step 3.$g_{hj}$.4.  
Let $n_{hj}$ be the length of a shortest path between $b_h$ and $b_j$ in $D-X_0$. We first show that the case
$n_{hj}>2$ reduces to the case $n_{hj}=2$ (recall that connectors and $B$-origins alternate on the paths).
Let $n_{hj}>2$ and recall that 3.$g_{hj}$.4 is the latest step used to decide that $b_h$ precedes $b_j$. 
Then exactly one pair $(b_u,b_w)$ of consecutive $B$-origins on the shortest path between $b_h$ and $b_j$ in $D-X_0$ 
is joined by a new 2-path built in step 3.$g_{hj}$.4: at least one
since step 3.$g_{hj}$.4 is needed, and at most one since otherwise two pairs  $(b_{u_1},b_{w_1})$ and $(b_{u_2},b_{w_2})$ in this order
on the path would satisfy  $b_{u_1}, b_{u_2}\in V(D')$ and $b_{w_1}, b_{w_2}\in V(D'')$ thus 
$b_{u_2}$ would precede $b_{w_1}$, whereas the viceversa already holds and thus $D$ would contain a circuit, a contradiction.
Consequently, the inductive hypothesis can be applied to paths between $b_h$ and $b_u$, 
respectively between $b_w$ and $b_j$, both built before step 3.$g_{hj}$.4 and allows to conclude 
(by (T) and a sorting of the increasing forcing sequences, as above) if we show that $\mathcal{(P.\rm{1})}$$\Rightarrow$ holds 
for $b_u$ and $b_w$, for which $n_{uw}=2$.

We thus study the case $n_{hj}=2$. Since step 3.$g_{hj}$ built the 2-length path between $b_h$ and $b_j$,
we have that $b_h\in V(D')$ and $b_j\in V(D'')$ in step 3.$g_{hj}$.4. Then consider the step 3.$g_{hj}$ and the current values 
$\Last_x$ of the vertices $b_x$ in $D-X_0$, {\em i.e.} the values $\Last_x$ computed in step 3.$g_{hj}$-$1$.  
With $b_h\in V(D')$ and $b_j\in V(D'')$ in step 3.$g_{hj}$.4 we deduce that $a_{g_{hj}}b_h\in E$, whereas $a_{g_{hj}}b_{j}\not\in E$,
implying that $g_{hj}\neq k_j$. But we know that:  (a) $b_j$ is adjacent to an $A$-origin $a_c$ with $c<g_{hj}$  (since $b_j$ belongs to $D$); 
and (b) $b_j$ has $\Last_j>g_{hj}$  since at the end of step $g_{hj}-1$ we necessarily have  $\Last_j>g_{hj}-1$ 
(otherwise $b_j$ would have been moved into $X_0$) and since $b_ja_{g_{hj}}\not\in E$.

In the case where  $a_{g_{hj}}$ is between $a_c$ and $a_{k_j}$,  (TB)  implies $b_h\precf b_j$ with support $a_{g_{hj}}$. 

In the case where $a_{g_{hj}}$ is between $a_{k_j}$ and  $a_{\Last_{j}}$, the latter one is an artificial $A$-origin with 
non-integer index, say $\Last_j=t+\frac{1}{2}$. We then have, as proved above,  $\Last_j= t+\frac{1}{2}>g_{hj}$, 
thus $t\geq g_{hj}$ (*). Moreover, since $\Last_j= t+\frac{1}{2}>g_{hj}$ at the beginning of step 3.$g_{hj}$,
by  Remark \ref{rem:predecessort} we deduce the existence of a 
predecessor $b_r$ of $b_j$ in $D-X_0$ at the beginning of step 3.$g_{hj}$, thus also at the end of the step 3.$g_{hj}-1$,
such that  $\Last_{r}=t$.
We deduce by inductive hypothesis that $b_{r}\precf b_j$, thus
$k_{r}\leq m_j$ (**). But since $\Last_{r}=t$ and $t$ is integer, we deduce $k_{r}=t$ and thus, using (*) and (**), 
$m_j\geq k_{r}=t\geq g_{hj}$. Now, recalling that we are considering the case where $a_{g_{hj}}$ is between 
$a_{k_j}$ and  $a_{\Last_{j}}$, we deduce $k_j<g_{hj}\leq m_j$. Then (FB') implies that
$b_h\precf b_j$ with support $a_{g_{hj}}$. The conclusion follows. 

\medskip

\noindent {\it Proof of} $\mathcal{(P.\rm{1})}$$\Leftarrow$:

Let $b_h\precf b_j$ be deduced from the order on $A$ only, using an increasing forcing sequence with supports $a_f\preceqf a_i$.
Let $S$ be such sequence. 
We show by induction on 
$k$ that when the $k$-th rule of $S$ decides that $b_u\precf b_w$ with support $a_f$,  then the algorithm decides 
in step 3.$f$.4 that $b_u$ precedes $b_w$.

When $k=1$, (TB) is applied with support $a_f$ to deduce that $b_u\precf b_w$. Then in step 3.$f$.4 we have that $b_u\in V(D')$,
since $a_fb_u\in E$, and $b_w\in V(D'')$ by invariant (I.2d) since $b_w$ has a neighbor before and a neighbor after $a_f$. Thus
$b_u$ precedes $b_w$ at the end of step 3.$f$.4.

When $k>1$ and the inductive hypothesis holds, when (TB) is applied then the reasoning above applies again. When (FB') is applied,
let $a_f$ be the support. Then $f\leq i$, $a_f\precf b_w$, deduced by the previous rules in $S$, $k_w<f$ (thus $a_fb_w\not\in E$) and 
$a_fb_u\in E$. Now,  $a_f\precf b_w$ implies by Proposition \ref{prop:situationsonA} one of the following cases: either 
(b) $a_f$ is a neighbor of  $b_x\precf b_w$, or (c) $a_f\precf a_g\precf b_w$, for appropriate $b_x$ and $a_g$, 
where the relations $\precf$ are 
previously deduced by $S$. In the former case, the inductive hypothesis for $b_x\precf b_w$ ensures that $b_x$ precedes $b_w$
in $D$ as a result of steps 3.$q$.4 with $q\leq f\leq i$ only, and the conclusion follows by noticing that in step 3.$f$.4 
we have $b_u, b_x\in V(D')$ and $b_w\in V(D'')$. The latter case reduces to the former one as follows. Choose $g$ as large 
as possible with the property that $a_f\precf a_g\precf b_w$ due to rules in $S$ applied before the current rule (TB'). Then, 
by Proposition \ref{prop:situationsonA} for $a_g\precf b_w$, one for the situations (a), (b), (c) must occur.
Situation (a) is not possible since $k_w<f$ and $g>f$, and situation (c) is not possible
by the choice of $a_g$. Then situation (b) occurs, and this is the former case.

\medskip

\noindent{\it Proof of}  $\mathcal{(P.\rm{2a})}$:

To prove that $\lfloor \Last_h\rfloor \leq m_h$ at the end of step 3.$i$.5, we note that initially $\Last_h=k_h\leq m_h$, and that $\Last_h$ increases to a new
value $t+\frac{1}{2}$ during the steps 3.$v$.5, with $v\leq i$, where $b_h$ gains a new predecessor $b_q$ with 
$\Last_q\in \{t,t+\frac{1}{2}\}$. However, when $\Last_q=t+\frac{1}{2}$ that means $\Last_q$ has been updated too, and 
by Remark \ref{rem:predecessort} we deduce 
the existence of a predecessor $b_r$ with $\Last_r=t$. But then  $\mathcal{(P.\rm{1})}$$\Rightarrow$ implies that
$b_r\precf b_h$ and thus $k_r\leq m_h$. Now, since $\Last_r=t$ and $t$ is an integer, we have that $t=k_r$ and thus
$\lfloor \Last_h\rfloor =t=k_r\leq m_h$.
\medskip

\noindent{\it Proof of}  $\mathcal{(P.\rm{2b})}$: 

By hypothesis, consider we are in step $i$ and that 
$i=\Last_h$. Then $\lfloor i\rfloor\leq m_h$. Assume by 
contradiction that the latter inequality is strict, implying that $i<m_h$.  By Proposition \ref{prop:situationsonA} 
for $a_{m_h}\precf b_h$,  one of the three situations (a), (b), (c) must occur.  In situation (a), $a_{m_h}b_h\in E$ and this
is not possible, since otherwise $m_h\leq k_h\leq  \lfloor \Last_h\rfloor<m_h$, a contradiction. Situation (c),
implying that $a_{m_h}\precf a_s\precf b_h$ for some $a_s$ cannot occur by the maximality of $m_h$. Then we are necessarily in case  
(b): $a_{m_h}$ is a neighbor of a $B$-origin $b_u$, but not of $b_h$, such that $b_u\precf b_h$. Thus $k_u=m_h>i$.
Then let $b_z$ be chosen such that $b_u\preceqf b_z\precf b_h$, $i<k_z$ and $b_z$ is as close as possible to $b_h$ 
with these properties. Such a $b_z$ exists, since $b_u$ belongs to the set of $B$-origins satisfying the abovementioned conditions.
We consider again the four cases in Proposition \ref{prop:3situations}, applied to $b_z\precf b_h$.

If $b_z$ is a basis of $b_h$ (case (1)), then its support $a_f$ in the rule (T) satisfies $f<k_h\leq  i$. In step
3.$f$.4, $b_z$ precedes $b_h$ and thus in step 3.$f$.5 $\Last_h$ is updated to a value  larger than or equal to $\Last_z$. But
$\Last_z\geq k_z> i$ so we cannot have $\Last_h= i$ in step 3.$i$.7.

If $b_z$ is a descendant of a basis $b_v$ of $b_h$ (case (2)), then using Proposition \ref{prop:status2} we deduce that
$i $ belongs to one of the intervals $I_1=[a_{u_0},a_{u_1}]$, $I_r=[a_{u_{r-1}}, a_{u_r}]$ for $2\leq r\leq g-1$ and
$I_g=[a_{u_{g-1}},a_{k_z}]$, since $k_h\leq  i <k_z$. Assume $i\in I_r$, $1\leq r\leq g$, such that 
$a_{i}$ is not the right endpoint of $I_r$ (choose $I_{r+1}$ if this happens). Consider the same notations as
in Proposition \ref{prop:status2}, with $b_{u_g}=b_z$.
Then in step 3.$u_{r-1}$.4, with $u_{r-1}\leq i$, we have $b_{u_r}\in V(D')$ whereas $b_h\in V(D'')$, thus $b_{u_r}$ precedes $b_h$.
Thus in step 3.$u_{r-1}$.5, $\Last_h$ is updated to a value larger than or equal to $\Last_{u_r}$, which is larger than $ i $ since
$u_r$ is adjacent to the right endpoint of $I_r$.
Thus we cannot have $\Last_h= i$ in step 3.$i$.7.

In cases (3) or (4), let $b_q$ be such that $b_q\precf b_h$ and $b_z$ is
either a basis or a descendant of a basis of $b_q$. Then $b_z\precf b_q\precf b_h$ and by the choice of $b_z$ we
must have $k_q\leq i $. Thus we can use the same reasoning as in (1) and (2) above, but this time for $ i , b_z$ and $b_q$,
to deduce that $\Last_q> i$ at the end of step 3.$ i $.5. Then it is sufficient to show that 
$b_h$ is a successor of $b_q$ in step 3.$ i$.5 in order to deduce again that $\Last_h$ should have been updated.
To this end, by
Corollary \ref{cor:inductive12} we deduce the existence of a sequence 
$t_1, t_2, \ldots, t_p$ of distinct integers such that $t_1=q, t_p=j$ and $b_{t_{r}}$ is either a basis or a descendant
of a basis of $b_{q_{r+1}}$, for $1\leq r\leq p-1$.
Then the corresponding forcing sequence for $b_{t_r}\precf b_{t_{r+1}}$ is increasing, and the last rule must use a support
$a_g\leq k_{t_r}$ since the support must be adjacent to $b_{t_r}$. It follows that all the supports $a_f$ used to deduce
$b_q\precf b_h$ satisfy $a_f\leq i $. In consequence,
$b_q\precf b_h$ is decided not later than step 3.$ i$, and thus $b_h$ is a successor of $b_q$ in step 3.$ i$.5.
\medskip

\noindent{\it Proof of}  $\mathcal{(P.\rm{2c})}$: 

When the algorithm is applied, at each moment  $\lfloor \Last_h\rfloor \leq m_h$, as proved in $\mathcal{(P.\rm{2a})}$. Since the number
of updates is limited and the sequence of values is strictly increasing, the algorithm will stop with a value maximum $\Last_h=j$ 
such that $\lfloor \Last_h\rfloor \leq m_h$. Note that $\Last_h$ cannot take successively both values $t$ (integer) and 
$t+\frac{1}{2}$, since then $t=k_h$ (the only integer value that $\Last_h$ can take), and the update of $\Last_h$ to
$t+\frac{1}{2}$ must be due to a predecessor $b_w$ with $\Last_w=t$, which is added in a step $u\leq t$. But then by
Lemma~\ref{lem:3} we deduce that $a_tb_h\not\in E$, which contradicts the equality $t=k_h$ we obtained above.

Then, in step 3.$j$, $b_h\in V(D-X_0)$ since $b_h$ cannot move into $X_0$ in a previous step. We 
then have $\Last_h=j$ in step 3.$j$.7, and  by affirmation $\mathcal{(P.\rm{2b})}$, 
we deduce that $\lfloor \Last_h\rfloor = m_h$. Thus $\lfloor j\rfloor=m_h$. As proved in the previous paragraph,
$\Last_h$ can take only one of the values $m_h$ and $m_{h}+\frac{1}{2}$, and thus so does $j$. 

\medskip

\noindent {\it Proof of} $\mathcal{(P.\rm{3})}$$\Rightarrow$: 

The hypothesis $b_h\not\precf b_j$ and $b_j\not\precf b_h$, and affirmation $\mathcal{(P.\rm{1})}$$\Rightarrow$ imply that
there exists a step 3.$i$ with $i\leq m$ such that $b_h\in V(F')$ in step 3.$i$.7, and either (a) $b_j$ belongs to $V(F'')$
but is not a successor of $b_h$ in $D-X_0$ in step 3.$i$.4, or (b) $b_j$  is added to $D$  in a subsequent step 3.$f$.7, $i<f\leq m$.
In both cases we deduce that $b_j$ has at least one neighbor $a_s$  with integer $s\leq m$.
Moreover, with $b_h \in V(F')$ we deduce $\Last_h=i$ in step 3.$i$.7 and affirmation $\mathcal{(P.\rm{2b})}$ implies 
$\lfloor i\rfloor=m_h$. Thus $m_h\leq m$. 

In case (a), $b_j\in V(F'')$ implies that $\Last_j>i$ in step 3.$i$.7 and we have two subcases. When 
$i$ is not integer, we have $m_h=\lfloor i\rfloor <i\leq \lfloor \Last_j\rfloor \leq m_j$, where the last inequality is obtained by 
$\mathcal{(P.\rm{2a})}$, and we are done. When
$i$ is integer, we have $m_h=\lfloor i\rfloor =i\leq \lfloor \Last_j\rfloor \leq m_j$. Then we can have $m_h=m_j$ only if 
$\Last_j=i+\frac{1}{2}$, which implies that
$a_ib_j\not\in E$ (Lemma \ref{lem:3}), whereas $a_ib_h\in E$, since $\Last_h=i$ is an integer  and this can hold only when $i=k_h$. Thus, in step 3.$i$.4,
$b_h\in V(D'), b_j\in V(D'')$ and thus $b_h$ precedes $b_j$ due only to step 3.$i$.4. By $\mathcal{(P.\rm{1})}$$\Rightarrow$
we deduce that $b_h\precf b_j$, a contradiction with the hypothesis. Thus $m_h<m_j$.

In case (b), $m_j\geq f>i\geq \lfloor i\rfloor =m_h$  and we are done.
\bigskip

\noindent {\it Proof of} $\mathcal{(P.\rm{3})}$$\Leftarrow$:

Consider now the case where $b_h\precw b_j$ with $m_h\leq m$.  By $\mathcal{(P.\rm{2c})}$,
$b_h$ belongs to $F'$ in step 3.$i$.7 with $i\in\{m_h, m_h+\frac{1}{2}\}$.  
Concerning $b_j$, we know that in step 3.$i$.7 either $b_j$ belongs to $F''$, and this happens when $b_j$ has a neighbor 
$a_s$ with $s\leq i$; or $b_j$ is added to $D$ in a later step 3.$s$ with $i<s\leq m$, and this happens when $b_j$ has a neighbor $a_s$
with $i<s\leq m$.  
In both cases, $m_j>m_h$ holds thus, by $\mathcal{(P.\rm{2b})}$, $b_j$ can move to $X_0$
only in step 3.$t$.7 with $t\in\{m_j, m_j+\frac{1}{2}\}$, thus with $t>i$. Then, in step 3.$i$.7, the placement of $b_h$ in $X_0$ 
ensures that $b_h$ becomes a predecessor of $b_j$ as soon as $b_j$ belongs to $D$, and not later than in  step 3.$s$, with $s\leq m$.

\medskip

\noindent {\it Proof of} $\mathcal{(P.\rm{4})}$$\Rightarrow$: 

Recall that $q$ is an integer. The $A$-origin $a_q$ appears in $D$ in step 3.$q$.7, as the source of $F'$ when $F'$ becomes a part of $X_0$. Then $q\leq m$.
Now,   $a_q$ precedes $b_j$ in the DAG $D$ obtained at the end of step 3.$m$ if and only if  
in step 3.$q$.7: either (i) $b_j$ belongs to $F'$, or (ii) it belongs to $F''$, or (iii) $b_j$ is inserted in the DAG $D$, and more
precisely in $D'$, in a further step. We consider each of them. In case (i), $a_qb_j\in E$ since $\Last_j=q=k_j$ (the only case where
$\Last_j$ is integer is when it has not been updated), and thus by (A) we have the conclusion; in case (ii),
$\Last_j>q$ implies by affirmation $\mathcal{(P.\rm{2b})}$ that $m_h=\lfloor \Last_j\rfloor\geq q$
and by the definition of $m_h$ that $a_q\precf b_j$; in case (iii), $b_j$ is
inserted in some step $s\leq m$, thus $a_sb_j\in E$. By rules (O) and (A) we have that $a_q\precf a_s\precf b_j$ and rule (T)
implies the conclusion.

\medskip

\noindent {\it Proof of} $\mathcal{(P.\rm{4})}$$\Leftarrow$:

By Proposition \ref{prop:situationsonA}, $a_q\precf b_j$ is implied by only three configurations. In configuration (a), 
$a_qb_j\in E$ and thus in step $q$ the vertex $b_j$ belongs to $D'$, and then either to $F'$ or to $F''$. In both cases
$a_q$ precedes $b_j$ at the end of step 3.$q$.7. In configuration (b), and if configuration (a) does not hold,
$a_qb_z\in E$ for some $b_z$ that satisfies $b_z\precf b_j$, and $a_qb_j\not\in E$. Then, in step $q$, $b_z$ belongs to $D'$, 
whereas $b_j$ may belong or not to $D''$. In case $b_j\in V(D'')$, we also have that $b_j\in V(F'')$ since 
$\Last_j>q$  in step 3.$q$.5 and thus also in step 3.$q$.7; therefore $a_q$, that belongs to $X_0$ at the end of step 3.$q$.7, 
precedes $b_j$. And if $b_j\not\in V(D'')$,
the neighbor $a_s$, $s\leq m$, of $b_j$ also satisfies $s>q$ and guarantees that $b_j$ is inserted in $D$ later. Since
$b_j$ is inserted in $D'$ and, once the step $q$ finished, $a_q$ belongs to $X_0$, we deduce
that $a_q$ precedes $b_j$.
\bigskip

\bproof {\bf (of Theorem \ref{thm:correction}.)}  We prove that $G$ is $A$-Stick iff Algorithm CSO stops at the end of 
step $|A|^+$, which is equal to $|A|+\frac{1}{2}$ if the artificial origin $a_{|A|+\frac{1}{2}}$ is created during the
algorithm, and to $|A|$ otherwise. To this end, we use the characterization of $A$-Stick graphs in Theorem \ref{thm:main}.
\medskip

\noindent{\it Proof of the backward direction.} If the algorithm stops at the end of the step $|A|^+$, then the final $D$ is a DAG 
containing all the vertices in $A\cup B$. Any circuit of  $\precf$ would induce a circuit in $D$, 
due to $\mathcal{(P.\rm{1})}$$\Leftarrow$, Theorem \ref{thm:increasing} and $\mathcal{(P.\rm{4})}$$\Leftarrow$, and such a circuit
does not exist in $D$. Thus $\precf$ is a partial order.
\medskip

\noindent {\it Proof of the forward direction.}
By hypothesis, we assume that $\precf$ has no circuits. Assume by contradiction that the
algorithm completely performs some step $m<|A|^+$, but stops in the next step $m^+$, which is necessarily integer.
Then in step 3.$m^+$.2 the DAG $D'$  is not compact, meaning that a  pair $(b_x,b_y)$ exists such that $First$ precedes $b_x$ which precedes 
$b_y$, and $b_x\in V(D'')$ whereas $b_y\in V(D')$. Then $b_x$ precedes $b_y$ in $D-X_0$ at the end of step $m$ too.
By  $\mathcal{(P.\rm{1})}$$\Rightarrow$, we deduce that $b_x\precf b_y$. We prove that $b_y\precf b_x$ also holds, thus 
contradicting the assumption that $\precf$ has no circuits.

With $(b_y,b_x)\in V(D')\times V(D'')$ in step 3.$m^+$.2 we deduce that $b_y$ is adjacent
to $a_{m^+}$,  whereas $b_x$ is not adjacent to  $a_{m^+}$. But $b_x$ satisfies: (a) is adjacent to an $A$-origin $a_c$ smaller than $a_{m^+}$ 
(since it belongs to $D$) and (b) has $\Last_x$ (computed in some step $v< m^+$) larger than $m^+-1$ and 
thus, since $a_{m^+}b_x\not\in E$, larger than $m^+$. That means  $a_{m^+}$ is either between $a_c$ and $a_{k_x}$ 
in which case (TB) implies $b_y\precf b_x$; or $a_{m^+}$ is between
$a_{k_x}$ and $a_{\Last_x}$, and more precisely between $a_{k_x}$ and $a_{m_x}$ due to $\mathcal{(P.\rm{2b})}$.  
In the latter case,  we deduce by (FB') that $b_y\precf b_x$. 

The correction of Algorithm CSO is now established. Moreover, affirmations 
$\mathcal{(P.\rm{1}), (P.\rm{3}),(P.\rm{4}) }$ ensure that the order output by the algorithm is a canonical order.
\eproof

\br
Note that several canonical orders may exist for an $A$-Stick graph $G$. Algorithm CSO outputs one of them, but 
records each of them in the final DAG $D$. So the algorithm is able to test whether a given $A$-Stick representation uses a canonical 
order or not. 
\label{rem:canonical}
\er

\subsection{Running time}\label{subsec:runningtime}

In this section we prove the following result.

\bthm
Algorithm CSO runs, with an appropriate implementation, in $O(|A|+|B|+|E|)$ time.
\ethm

\bproof
The invariants (I.1), (I.3) and (I.4) allow us to deduce that $|V(D)|=O(|A|+|B|)$, and $|E(D)|=O(|A|+|B|)$, 
at each moment in the algorithm. The most difficult step to
implement is step 3, that we explain in detail. For a non-integer $i$, we define $N(a_i)$ to be the set of 
vertices $b_y$ for which the final value of $Last_y$ is $i$. Then note that, for each small bubble $Y_i$,
$V(Y_i)\cap B\subseteq N(a_i)$.

We also define the {\em right-connector set} of a compact induced sDAG $Q$ of $D$ as the set: 
\medskip

$C(Q)=\{c\, |\, c\in V(Q)\, \hbox{or}\, \hbox{all the in-neighbors of}\, c\, \hbox{belong to}\, V(Q),\, \hbox{and}\, c\, \hbox{has an out-neighbor in}\, D-Q\}$
\smallskip

\noindent and the set $\mathcal{F}(Q)$ of {\em frontiers} of the induced sDAG $Q$ as

$$\Front(Q)=\{b_j\in B\, |\, b_j\not\in V(Q), \exists c\in C(Q): cb_j\in E(Q)\}$$

Then the steps are implemented as follows. 
With the aim of speeding up the treatment, the DAG $D$ is implemented with opposite arcs, although only the direction from
left to right gives the order of vertices we are interested in. Note that when $N(a_i)=\emptyset$, the running time of step 3 is in $O(1)$.

\begin{enumerate}[leftmargin=1.1cm]
  \item[3.$i$.1:] We use an array of pointers to the vertices $b_j$ of $D$ (the pointer is null if $b_j\not\in V(D)$).
 \item[3.$i$.2:] This step returns either a pair $(b_x,b_y)$ of $B$-origins showing that $D'$ is not compact, or the 3-tuple
 $(D',s_t,C)$ such that $s_t$ is the rightmost strong connector in $D'$ and $C=C(D')$. 
 We identify $D'$, its targets and test the possible existence of a pair $(b_x,b_y)$ contradicting the compactness using
 disjoint depth-2 backward traversals from each vertex in $N^D(a_i)$. Then, $C$ and $s_t$ are computed in another traversal
 starting with $First$. The two traversals (backward and forward) visit a number of vertices 
 ($B$-origins and connectors) proportional to $|N^D(a_i)|$.
 The running time of this step, as well as the cardinalities of $V(D')$ and $C$, are thus in $O(|N^D(a_i)|)$.
 \item[3.$i$.3:] The update of the DAG $D'$ is linear in $|N^{\overline{D}}(a_i)|$. Now, $|V(D')|$ and $|E(D')|$ are in $O(|N(a_i)|)$.
 \item[3.$i$.4:] We use $(D',s_t,C)$ computed above. Since $D'$ is compact, all the targets of $D'$ belong to $X_{s_t}$ and the 
 frontiers in $\Front(D')$ either belong altogether to $X_{s_t}$, or are the out-neighbors of $s_u$, the closing connector of $X_{s_t}$.
 The frontiers are then respectively called {\em close} and  {\em distant}, and we denote $\Front^c(D')$ respectively $\Front^d(D')$
 the sets regrouping each type of frontiers. Thus $\Front(D')=\Front^c(D')\cup \Front^d(D')$ with either $\Front^d(D')=\emptyset$
 or $\Front^c(D')=\emptyset$. 
 
 The latter case corresponds to a step 3.$i$.4 where there is no $B$-origin in $V(X_t)-V(D')$, so the bubble of $D'$ is already closed. 
It takes $O(1)$ to identify this case, characterized by $|C|=1$ such that the unique element of $C$ is a strong
connector different from $s_t$. In the former case, create a new, simple, connector $s'_t$, with the same in-neighbors as $s_t$ 
but only with its out-neighbors from $D'$. This takes $O(|V(D')|)$, i.e. $O(|N(a_i)|)$, time if 
$s_t\neq First$ since all the in-neighbors of $s_t$ are in $D'$, and takes $O(|N^-(First)|+|N(a_i)|)$ time 
if $s_t=First$. Then remove, with the same running time, all the arcs towards $s_t$ and 
the arcs from $s_t$ to its out-neighbors from $D'$ (but keep the arcs from $s_t$ to its out-neighbors, since they are frontiers). 
We have to make $s_t$ be the new strong connector connecting the targets in $D'$
and all the frontiers of $D'$, all of which are close but some of which are not yet known. Then discover all the $B$-origins $b_h\in \Front^c(D')\setminus N^+(s_t)$ using the 
connectors in $C\setminus \{s_t\}$  and replace the arc ingoing to $b_h$ by the arc $s_tb_h$.  This takes 
$O(|\Front^c(D')\setminus N^+(s_t)|)$ time.
It remains to add arcs from all the targets of $D'$ to $s_t$, which is done in $O(|N(a_i)|)$ time. 
The other operations in this step need $O(|N(a_i)|)$ time. In conclusion, the running time of this step is 
in $O(|N^-(First)|+  |N(a_i)|+|\Front^c(D')\setminus N^+(s_t)|)$ time when the frontiers are close, and in 
$O(1)$ when the frontiers are distant.
 
\item[3.$i$.5:] The update of $\Last_j$ for $b_j\in V(D-X_0)$ indicated in Algorithm CSO is too long. 
It is therefore replaced by $(1)$ the update of $\Last_j$ for the vertices $b_j$ in $V(D')$ only, 
as described in step 3.$i$.5, and $(2)$ an 
update in step 3.$i$.6, that concerns only the vertices $b_z$ with $\Last_z=i$ when $i$ is not integer.
%
%
%
%
With this approach, for each $b_j$
 its value $\Last_j$ is considered for a possible update only in the steps where $b_j$ belongs to $D'$ (update $(1)$), 
 and - finally - in the step 3.$i$ such that $\Last_j=i$ (update $(2)$ described in step 3.$i$.6). 
 In the intermediate steps where $b_j$ is in neither of these cases,
 the possible updates of $\Last_j$ are due to predecessors $b_w$ with current value $\Last_w>\Last_j$. Only the largest such value,
 that is $\Last_w=i$, is worth being recorded for $\Last_j$. This can be done in step 3.$i$, since $b_w$ belongs to $D-X_0$
 in step 3.$i$.5 (due to $\Last_w=i$) and still precedes $b_j$ (Remark \ref{rem:alwaysprecede}).

\item[3.$i$.6:] Note that, by the definition of $\Last_j$, $F'$ is always compact.
When $i$ is an integer, the vertices $b_j$ with $\Last_j=i$ belong to $D'$ and thus their $\Last_j$
is necessarily updated in step 3.$i$.5. Computing $F'$ in this step is done similarly to the computation of $D'$ in 
step 3.$i$.2, with the same running time.  

When $i$ is not an integer, some $B$-origins $b_j$ from $D-X_0$ whose $\Last_j$ should be equal to 
$i$ may have a current value $\Last_j$ less than $i$, if the necessary update has not been performed yet. (In any case, the
updated value of $\Last_j$ must be at least $i$, otherwise $b_j$ would have already been placed in $X_0$.)
The $B$-origins $b_j$ that must be put in $F'$ in step 3.$i$.6 are exactly those with the property that $\Last_j\leq i$ 
and such that no possible update to a value larger than $i$ is possible.
In order to find them, an easy approach is to
start with $First$ and to perform a depth-first traversal of $D$ which stops its progression along a branch
 as soon as either a vertex $b_u$ with $\Last_u>i$ is encountered, or  
 a connector having a predecessor $b_u$ with $\Last_u>i$ is encountered. For the latter test, we only need $O(1)$ time to compare 
 - each time the traversal reaches the connector - the total number of in-neighbors of the connector to the number of its visited in-neighbors 
 $b_z$ (which are exactly those satisfying $\Last_z\leq i$).
 Then $F'$ is the subgraph of $D$ induced by the vertices $b_j$ with $\Last_j\leq i$ encountered during the traversal,
 and their previous connectors (among which $First$, the source of $F'$). The rightmost strong connector $s_q$ in $F'$, the targets of
 $F'$
 and the right-connector set $C$ of $F'$ are also computed during the traversal.
 
 However, this approach takes $O(|V(F')|+|\Front(F')|)$ time, whether the frontiers are close or distant, since all the frontiers are 
 visited. We reduce this time to $O(|V(F')|+|\Front^c(F')\setminus N^{+}(s_q)|$ time, where $s_q$ is the rightmost strong connector
 in $F'$, as follows. Recall that $i$ is not integer, {\em i.e.} $i=h+\frac{1}{2}$, where $h$ is an integer.
 
 According to our approach, the out-neighbors of a connector $s_p$ from $D-X_0$ are visited only if no predecessor $b_x$ of $s_p$ exists in
 $D-X_0$ such that $\Last_x>i$. Then we may partition $N^+(s_p)$ as $N^{+}(s_p)=N^{+,int}(s_p)\cup N^{+,frac}(s_p)$, where
 \medskip
 
 \hspace*{1cm} $N^{+,int}(s_p)=\{b_f\in N^{+}(s_p)\, |\, \Last_f\, \hbox{is integer and}\, \Last_f>i\}$
 
  \hspace*{1cm}  $N^{+,frac}(s_p)=\{b_f\in N^{+}(s_p)\, |\, \Last_f\, \hbox{is not integer and }\Last_f\leq i \}$
 \medskip
 
 \noindent We first explain why  $N^{+}(s_p)$ cannot contain $B$-origins $b_f$ with integer $\Last_f$ such that $\Last_f< i$. Assume by
 contradiction that such a $B$-origin $b_f$ exists and let $t=\Last_f$, in the current step 3.$i$.6. Then $t=k_f$, since this is
 the only possible integer value of $\Last_f$. In step 3.$t$.5 two cases occur: either $\Last_f$ remains equal to $t$, in which case 
 in step 3.$t$.7 $b_f$ is moved to $X_0$, and this contradicts the hypothesis that $b_f\in N^{+}(s_p)$, since $s_p\in V(D-X_0)$;
 or $\Last_f$ is updated, in which case $\Last_f$ is no longer an integer in step 3.$i$, with $i>t$, another contradiction.
  
 Moreover, $N^{+}(s_p)$ cannot contain $B$-origins with  $\Last_f=t+\frac{1}{2}>i$, where $t$ is an integer. In the contrary case, 
  $\Last_f=t+\frac{1}{2}>i=h+\frac{1}{2}$  implies $t>h$ and, given that both $t$ and $h$ are integer values, that $t>i$.
  By Remark \ref{rem:predecessort}, $b_f$ has, in the current $D-X_0$, a predecessor $b_r$ with $\Last_r=t$. Since $b_f\in N^+(s_p)$,
  $b_r$ must precede $s_p$, and this contradicts the assumption we made that no predecessor $b_x$ of $s_p$ exists in
 $D-X_0$ such that $\Last_x>i$.
 \newpage
 
 Now, the vertices in $N^{+,int}(s_p)$ are necessarily frontiers, whereas those in $N^{+,frac}(s_p)$ necessarily belong to $F'$.
 Then the rightmost strong  connector $s_q$ of $F'$ is detected as follows: it is the first  strong connector $s_p$ encountered
 during the traversal for which either $N^{+,int}(s_p)\neq \emptyset$, or there is a connector $s_r$ belonging to $X_{s_p}$ such that
 $N^{+,int}(s_r)\neq \emptyset$. 
 The frontiers are all distant iff we are in the first case and $N^{+}(s_p)=N^{+,int}(s_p)$,
 and they are all close in the contrary case.
 
 Consequently, in order to identify the frontiers it is sufficient to represent the list of out-neighbors of the connectors 
 as two double linked sublists, containing respectively the vertices $b_j$ with integer and non-integer value $\Last_j$, in order to avoid visiting the
 distant frontiers and the close frontiers in $N(s_q)$. An update of $\Last_j$ may move $b_j$ from the first list to the second one,
 which is realized in $O(1)$ time using the array in step 3.$i$.1.

\item[3.$i$.7:] This operation is similar to that in step 3.$i$.4, and thus needs $O(|N^-(First)|+
 |N(a_i)|+|\Front^c(F')\setminus N^+(s_q)|)$ time when the frontiers are close, and in 
$O(1)$ otherwise; and this, whether $i$ is integer or not.

\end{enumerate}
\bigskip

{\bf Running time of Step 3.}  For a fixed value of $i$ integer or not, steps 3.$i$.1 to 3.$i$.7 require at most 
$O(1+|N(a_i)|+|N^-(First)|+|\Front^c(D')\setminus N^+(s_t)|+|\Front^c(F')\setminus N^+(s_q)|$ time, where
$1$ stands for the case where $N(a_i)=\emptyset$, the sets  $N^-(First)$, $\Front^c(D')\setminus  N^+(s_t)$ 
and $\Front^c(F')\setminus N^+(s_q)$ are those in step $i$, and $N^+(s_t)$ (respectively $N^+(s_q)$) is the 
out-neighborhood of $s_t$ (respectively $s_q$) before 
closing the bubble of $D'$ (respectively $F'$). After the bubble closing, every vertex in $\Front^c(D')\setminus N^+(s_t)$
(respectively of $\Front^c(F')\setminus N^+(s_q)$) becomes a {\em new} neighbor of $s_t$ (of $s_q$ respectively).

Now, $N^{-}(First)\subseteq V(Y_{prec(i)})\subseteq N(a_{prec(i)})\cup \{a_{prec(i)}\}$, as noticed at the beginning of the proof,  
where $prec(i)=i-\frac{1}{2}$ if $a_{i-\frac{1}{2}}$ exists, and $prec(i)=i-1$ otherwise. 
Moreover, define $\Front^{c,1}_{i,t}$ to be 
$\Front^c(D')\setminus N^+(s_t)$ if $s_t$ is the rightmost strong connector in $D'$ in step 3.$i$.4, 
and the empty set otherwise. We notice that:

(a) $\Front^{c,1}_{i,t}\subseteq N(a_t)$, and 

(b) $\Front^{c,1}_{i,t}\cap \Front^{c,1}_{j,t}=\emptyset$. 

\noindent Affirmation (a) follows from the remark that the close frontiers belong to $X_{s_t}$, and $X_{s_t}\subseteq N(a_t)$, 
by invariant (I.2). To show affirmation (b), assume w.l.o.g. that $i<j$, so 
that at the end of step 3.$i$.4 each $b_x\in \Front^{c,1}_{i,t}$ is a new out-neighbor of $s_t$. Then $b_x$ cannot be
a new out-neighbor of $s_t$ at the end of step $j>i$, since then the path between $s_t$ and $b_x$ in $D-X_0$ should be extended in
some step $3.l$ with $i<l<j$ before it is reduced again to an arc in step 3.$j$.4; but there is no instruction in the 
algorithm that extends a path between two $B$-origins in $D-X_0$. Thus (b) is proved. 

The same reasoning holds for  $\Front^{c,2}_{i,q}$, defined as $\Front^c(F')\setminus N^{+}(s_q)$ if $s_q$ is the rightmost strong connector in $F'$ in 
step 3.$i$.7, and the empty set otherwise. We then have (a$'$) $\Front^{c,\rm 2}_{i,q}\subseteq N(a_q)$ and (b$'$) 
$\Front^{c,\rm 2}_{i,q}\cap \Front^{\rm 2}_{j,q}=\emptyset$. Moreover,
we also have that, with $q=t$,  (b$''$) $\Front^{c,\rm 2}_{i,t}\cap \Front^{c,1}_{j,t}=\emptyset$ since the treatment 
of each of them consists in the irreversible process of transforming a vertex which is a successor but not an out-neighbor of $s_t$ 
into an out-neighbor of $s_t$. 

By (a), (a$'$), (b), (b$'$) and (b$''$) we deduce that for each integer $t$, $1\leq t\leq |A|$, we have 
$\Sigma_{\hbox{step}\, i} (|\Front^{c,1}_{i,t}|+|\Front^{c,\rm 2}_{i,t})|)=O(|N(a_t)|)$.

With the notations $t_i$ and respectively $q_i$ for the integer indices $t$ and $q$ used in steps 3.$i$.4
and 3.$i$.7, the running time $R(|A|,|B|,|E|)$ of step 3 is evaluated as follows. Here, step 3.$i$ is shortened as step $i$. 

\begin{align*}
R(|A|,|B|,|E|)&=
O(\Sigma_{\hbox{step}\, i} (1+|N(a_i)|+|N(a_{prec(i)})|+|\Front^{c,1}_{i,t_i}|+|\Front^{c,\rm 2}_{i,q_i}|))\\
&=O(|A|+ {\rm 2}\Sigma_{\hbox{step}\, i}  |N(a_i)|+ \Sigma_{\hbox{step}\, i} (|\Front^{c,1}_{i,t_i}|+|\Front^{c,\rm 2}_{i,q_i}|))\\
&=O(|A|+{\rm 2}\Sigma_{\hbox{step}\, i}  |N(a_i)|+ \Sigma_{t\, \hbox{integer}} \Sigma_{\hbox{step}\, i} (|\Front^{c,1}_{i,t}|+
|\Front^{c,\rm 2}_{i,t}|))\\
&=O(|A|+2\Sigma_{\hbox{step}\, i} |N(a_i)|+\Sigma_{t\, \hbox{integer}}|N(a_t)|)\\
&\leq O(|A|+3\Sigma_{\hbox{step}\, i} |N(a_i)|)\\
&= O(|A|+3\Sigma_{\hbox{integer step}\, i} |N(a_i)|+3\Sigma_{\hbox{non-integer step}\, i} |N(a_i)|)=\\
&=O(|A|+3|E|+3|B|)=O(|A|+|B|+|E|).
\end{align*}

The affirmation that $\Sigma_{\hbox{non-integer step}\, i} |N(a_i)|$ is in $O(|B|)$, used above, is due to the definition of 
$N(a_i)$ when $i$ is non-integer: it contains the vertices $b_x$ whose final value $Last_x$ is $i$. A vertex can have only one
final value $Last_x$.

\bigskip

\noindent{\bf Running time of Step 4.}
Finally, step 4 uses a traversal of $D$ starting with $a_1$, which visits all the paths inside a small bubble before moving to the
next small bubble. The vertices labeled with artificial origins are not output.  The running time of this traversal is 
in $O(|A|+|B|)$.\eproofs

\section{Stick representations of minimum length}\label{sec:length}

In this section, we say that a Stick representation is {\em steady} if the distance along the ground line between two consecutive 
origins is equal to 1. A Stick representation is a {\em shortest Stick representation} if the tip of each $B$-segment $B_x$ is
placed on the highest segment it must intersect, {\em i.e.} segment $A_{1_x}$, and similarly for the $A$-segments. 
The {\em length} of a $B$-segment $B_x$ is then the number of origins on the ground line situated strictly between its origin $b_x$
and $a_{1_x}$, plus 1. The definition is similar for the length of an $A$-segment. An $A$-Stick representation is {\em canonical}
if it uses a canonical order, and it is both steady and shortest.
By convention, in this section we assume that all the representations are steady and shortest.

We define the {\em length} of a Stick representation $R$ of $G=(A\cup B, E)$, denoted $length(R)$,  as the total length of its segments. 
Consider the problem:
\bigskip

\noindent{\sc MinLength Stick}

\noindent{\bf Input:} A Stick graph $G=(A\cup B, E)$.

\noindent{\bf Output:} Find a Stick representation of $G$ with minimum length.
\bigskip

The variants {\sc MinLength $A$-Stick} and {\sc MinLength $AB$-Stick} are obtained when the order on $A$, respectively the order on $A$ and
(separately) the order on $B$, are given. 

We consider below the two variants above, and mainly the most difficult one,  {\sc MinLength $A$-Stick}, with the aim of 
giving partial results related to canonical orders. 

{\sc MinLength $AB$-Stick} is an easy problem.
Each $AB$-Stick representation may be transformed into a left-optimal Stick representation by successively moving each $B$-origin in
increasing order towards its leftmost possible place. The resulting steady order is necessarily of minimum length: in each different 
placement that keeps the order of the $B$-origins, at least one $B$-origin is not left-optimal, and this increases the length of its associated segment.

We thus focus on {\sc MinLength $A$-Stick}. In this case, we can find examples of $A$-Stick graphs for which
non canonical $A$-Stick representation is of minimum length. 

\bex
The graph $G$ with vertex sets $A=\{a_1, a_2, a_3, a_4, a_5, a_6\}$ and $B=\{b_1,b_2\}$, and edges given by $N(b_1)=\{a_5\}$
and $N(b_2)=A$ has a unique canonical representation $C$ with $b_1$ between $a_5$ and $a_6$, and $b_2$ after $a_6$. However, moving $b_1$ after 
$b_2$ yields a Stick representation $R$ satisfying $length(R)=length(C)-3$.
\eex

We devote the remaining of this section to the identification of a sufficient condition for a graph $G=(A\cup B, E)$ to have a canonical 
$A$-Stick representation of minimum length. Given two $B$-origins $b_s$ and $b_t$, we say that $N(b_s)$ and $N(b_t)$ {\em strictly overlap} if 
$N(b_s)\not\subseteq N(b_t)$ and $N(b_s)\not\subseteq N(b_t)$. A graph $G$ such that, for any pair of $B$-vertices, their
neighborhoods strictly overlap is called an {\em N-overlap graph}.

Our main result in this section is given below.

\bprop
Let $G=(A\cup B,E)$ be an $A$-Stick N-overlap graph. 
Then there is a unique canonical  order for $G$, and the resulting canonical $A$-Stick representation is of minimum length.
\label{prop:canminlength}
\eprop

We need to prove several preliminary results, the first of which is for general $A$-Stick graphs.

\blem
Let $G=(A\cup B,E)$ be an $A$-Stick graph,  $\prec^c$ be a canonical order of it and $b_t, b_s$ be two vertices such that
$b_t\prec^c b_s$ and $1_s\leq 1_t$. Then $b_t\precf b_s$ or $N(b_t)\subseteq N(b_s)$. 
\label{lem:choix}
\elem

\bproof
Assume that $N(b_t)\not\subseteq N(b_s)$, and let $a_v\in N(b_t)\setminus N(b_s)$.  Then $v\leq k_s\leq m_s$.
Given that $\prec^c$ is canonical and $b_t\prec^c b_s$, we deduce that either $b_t\precf b_s$, and we are done;
or $b_t\precw b_s$ and in this case $b_t\not\prec^c b_s$ and $m_s\geq m_t$. In the latter situation,  in case $k_s>v$ (TB) and Remark \ref{rem:successive} 
may be applied to $b_t, a_v, a_{1s}, a_{k_s}$ to deduce that $b_t\precf b_s$, a contradiction; and in
case $k_s<v$, we have that $k_s<v\leq m_t\leq m_s$ and $a_vb_t\in E$, thus (FB') may be applied to deduce again that $b_t\precf b_s$.
\eproofs

All the results below assume without recalling it that  $G$ is  an $A$-Stick N-overlap graph. They prepare a reasoning by contradiction.
Assuming a given canonical order does not result in a minimum length $A$-Stick representation, they help us to know
which pairs $b_t, b_s$ can be swapped (Lemma \ref{lem:Q1}), in which cases the swap decreases the length of the representation 
(Lemma \ref{lem:Q}) and under which conditions the existence of such a swap is ensured (Lemma \ref{lem:swapornot}). These
tools are then used to prove Proposition \ref{prop:canminlength}.

\blem
Let $\prec^c$ be a canonical order of $G$. If $b_t\prec^c b_s$ then $b_t\precf b_s$ or $1_t<1_s$.  
\label{lem:Q1}
\elem

\bproof Assume the contrary
holds. Then by Lemma \ref{lem:choix} we deduce that $N(b_t)\subseteq N(b_s)$, which contradicts the hypothesis. 
\eproofs

Let $\prec$ be an $A$-Stick order of $G$, and $b_t\prec b_s$ be two $B$-origins that are consecutive with respect to $\prec$.
Since $N(b_t)\not\subseteq N(b_s)$, let $c$ be the minimum index such that $a_cb_t\in E$ and $a_cb_s\not\in E$. Now, let 
an {\em alternate chain} be a maximal sequence of alternating $A$-origins and $B$-origins $a_{x_1}:=a_c, b_{y_1},   a_{x_2}, b_{y_2}, \ldots, 
a_{x_l}, b_{x_l}, a_{x_{l+1}}$ such that for each $g$ with $1\leq g\leq l$: (a) $a_{x_g}b_{x_g}\in E$ and $a_{x_{g+1}}b_{x_g}\in E$;
(b) $x_{g+1}<x_{g}$; (c) $a_{x_{g}}b_s\not\in E$ (including for $g=l+1$). We call the {\em limit} of $b_s$ and $b_t$ the
index 
\medskip

$v_{st}=\min\{v\, |\, \hbox{there is an alternate chain such that}\, a_{l+1}=a_v\}$. 
\medskip

Note that if $l=0$, 
then $v_{st}=c$.
Furthermore, call a {\em swap} of $b_t$ and $b_s$ the operation that moves $b_s$ {\em immediately} before the limit $a_{v_{st}}$. 
A swap is {\em good} if   
 the resulting representation is $A$-Stick and it has strictly smaller length than the initial one, 
 assuming both representations are steady and their segments are as short as possible.

\blem
Let $\prec$ be an $A$-Stick order of $G$, and let  $b_t\prec b_s$ be two $B$-origins that are consecutive with respect to $\prec$.
The swap of $b_t$ and $b_s$ is good if and only if $1_s\leq 1_t$ and $b_t\not\precf b_s$.
\label{lem:Q}
\elem

\bproof
In the forward direction, $b_t\not\precf b_s$ is necessary since the resulting representation is $A$-Stick too. Assuming
by contradiction that $1_s>1_t$ and considering that $N(b_s)\not\subseteq N(b_t)$, we deduce that each $a_q\in N(b_s)\setminus N(b_t)$
(and there exists at least one) satisfies $b_t\prec a_{q}\prec b_s$. 
Then $b_s$ cannot move before $b_t$ since we would have $a_{q}b_s\not\in E$, a contradiction.

Conversely, since $N(b_t)\not\subseteq N(b_s)$, there exists $a_c$ such that $a_cb_t\in E$ and $a_cb_s\not\in E$. We choose 
$c$ as small as possible with this property. Then $k_s<c$, otherwise with $a_{1_s}, a_c, a_{k_s}$, we deduce by (TB) that 
$b_t\precf b_s$. Let $v_{st}$ be the limit of $b_s$ and $b_t$, and notice that $k_s<v_{st}\leq c$. To see this, we show
that for each $b_q$ belonging to an alternate chain and such that  $a_{k_s}\prec b_q$, we have $c_q>k_s$, where $c_q$
is minimum such that $a_{c_q}\in N(b_q)\setminus N(b_s)$. Indeed, $c_q\leq 1_s$ is not possible 
since then $N(b_s)\subseteq N(b_q)$, which contradicts the hypothesis;
and $1_s<c_q<k_s$ implies by (TB) that $b_q\precf b_s$, thus the chain containing $b_q$ is 
a sequence of descendants of $b_q$ with respect to $b_s$, which finally implies by (FB) that $b_t\precf b_s$, a contradiction.
Thus $c_q>k_s$ for each $B$-origin $b_q$ belonging to an alternate chain and such that $a_{k_s}\prec b_q$. As the
$A$-origin $a_u$ following $b_q$ in the chain satisfies the conditions (a) and (c) in the definition of an alternate chain,
and $c_q$ is the minimum index such that $a_{c_q}$ satisfies conditions (a) and (c), we deduce that $k_s<c_q\leq u\leq c$.
Thus $k_s<v_{st}\leq c$. 

Now, we first note that there is no possible place for $b_s$ between $a_{v_{st}}$ and $b_t$: in that case $b_s$ would intersect
one of the $A$-segments whose origins belong to the chain ending in $A_{v_{st}}$, a contradiction since $k_s<v_{st}$. 
In order to prove that $b_s$ may be placed {\em immediately} before $a_{v_{st}}$ without modifying the required segment intersections, 
we have to show that there is no $a_y$, with $y<v_{st}$, such that $a_yb_s\not\in E$ but the segments $A_y, B_s$ intersect. 
In the contrary case, since the segments are as short as possible, there exists a segment $B_z$ that intersects $A_y$, 
with $a_{v_{st}}<b_z$.
But this contradicts the choice of $a_{v_{st}}$, since there is an alternate chain containing $b_z$ and $a_y$ whose last element
$a_y$ satisfies $y<v_{st}$. 

We deduce that when $b_s$ is placed immediately before $a_{v_{st}}$, the resulting representation is $A$-Stick. It remains to show that
it has smaller length than the initial $A$-Stick representation. Let $k$ be the number of origins between $a_{v_{st}}$ 
(included) and $b_s$ (not included). The segments whose lengths are concerned by the swap, and their length modification 
(in parenthesis) are:  $B_s$ ($-k$);  $B_q$, for each $b_q$ between $a_s$ and $b_t$ included 
(at most +1 for each $q$); $A_y$ with $y>v_{st}$ (0);  $A_y$ with $y<v_{st}$ (between $-1$ and $-k$ if their tip 
belonged initially to $b_s$, $0$ otherwise).  The balance of these changes is always strictly negative, so the swap is good. \eproofs

\blem
Let $\prec$ be an $A$-Stick order of $G$, and let  $b_t\prec b_s$ be two $B$-origins such that $1_s\leq 1_t$ and $b_t\not\precf b_s$. 
Then one of the following holds:

a) either a good swap exists, between consecutive $B$-origins $b_{x}$ and $b_y$ such that $b_t\preceq b_x\prec b_y\preceq b_s$. 

b) or there exists $b_q$ with $b_t\prec b_q\prec b_s$ such that $b_t\precf b_q$ and $1_q<1_s$.
\label{lem:swapornot}
\elem

\bproof
The proof is by induction on the number $d_{ts}$ of origins between $b_t$ and $b_s$ with respect to $\prec$.

When $d_{ts}=0$, Lemma \ref{lem:Q} implies that the swap of $b_t$ and $b_s$ is good. When $d_{ts}>0$, and assuming the lemma holds for
smaller values of $d$, let $(b_{t'}, b_{s'})$ be the rightmost pair of consecutive $B$-origins between 
$b_t$ and $b_s$ such that $1_{s'}\leq 1_{t'}$. Then either $s'\neq s$ and $1_{s'}< 1_s$, or $s'=s$.

In the case where $s'\neq s$ and $1_{s'}< 1_s$, if $b_t\precf b_{s'}$ then b) is true with $q=s'$. If $b_t\not\precf b_{s'}$, from 
$1_{s'}< 1_s\leq 1_t$  we deduce by the inductive hypothesis, since $d_{ts'}<d_{ts}$, that either a good swap exists and a) holds, 
or that some $b_q$ with  
$b_t\precf b_q$ and $1_q<1_{s'}$ exists between $b_t$ and $b_{s'}$. In the latter case, we have that $1_q<1_{s'}<1_s$ and thus
b) is also true for $b_t$ and $b_s$.

In the case where $s'=s$, we have $1_{t'}\geq 1_s$ and if $b_{t'}\not\precf b_s$ then by Lemma \ref{lem:Q} we deduce that a) holds.
If $b_{t'}\precf b_s$, then let $b_{t''}$ be the $B$-origin closest to $b_s$ such that $b_{t''}\prec b_s$ and $b_{t''}\not\precf b_s$.
Then $b_t\preceq b_{t''}$, since $b_t$ satisfies the abovementioned conditions. Now, the order between $1_{t''}$ and $1_s$ creates
two subcases:

\begin{itemize}
 \item When $1_{t''}<1_s$, we deduce $1_{t''}<1_s\leq 1_t$ and either $b_t\not\precf b_{t''}$ in which case the inductive hypothesis
for $b_t$ and $b_{t''}$ allows to conclude; or $b_t\precf b_{t''}$ and thus b) holds with $q=t''$. 

\item When $1_{t''}\geq 1_s$, then we need to distinguish two cases again. If $t''\neq t$, then since $b_{t''}\not\precf b_s$ by the choice of 
$b_{t''}$ we deduce by inductive hypothesis for $b_{t''}$ and $b_s$ that a good swap exists. The situation b) cannot occur, since $b_q$ would contradict the choice of
$b_{t''}$. Finally, if $t''=t$, let $b_{v}$ be the $B$-origin situated immediately to the right of $b_t$. 
Then $b_{v}\precf b_s$, thus $b_t\not\precf b_v$. Now, $1_v\leq 1_{t}$ implies a) by Lemma~\ref{lem:Q} and we are done. We show that opposite case cannot occur. 
By contradiction, if $1_v>1_{t}$,  let $a_c\in N(b_t)\setminus N(b_s)$. Then $k_s<c$, otherwise $1_s<c<k_s$ and (TB) implies $b_t\precf b_s$,
a contradiction. Also, $k_{v}<c$ otherwise $k_s <c\leq k_{v}\leq m_s$, since $b_v\precf b_s$. By rule (FB') 
we deduce $b_t\precf b_s$ again. But then $1_t<1_v\leq k_v<c\leq k_t$, and $b_t\prec b_v$ implies that $N(b_v)\subseteq N(b_t)$, a contradiction.
\end{itemize}

The proof of the lemma is now complete. \eproofs

We are now able to prove Proposition \ref{prop:canminlength}.
\bigskip

\bproof {\bf (of Proposition \ref{prop:canminlength}).}
Lemma \ref{lem:Q1} implies $\prec^c$ is unique, since otherwise two elements $b_s,b_t$ that are not in the same order in 
two different canonical orders imply both  $1_t<1_s$ and $1_s<1_t$, a contradiction.

Now, assume - again by contradiction - that the canonical $A$-Stick representation is not optimal with respect to the minimum length, 
and let $\prec^{opt}$ be an optimal $A$-Stick representation of $G$. Then let $b_h$ and $b_r$ be two $B$-origins such that
$b_h\prec^c b_r$ and $b_r\prec^{opt} b_h$. We deduce that $b_h\not\precf b_r$ and $b_r\not\precf b_h$. Thus, by Lemma \ref{lem:Q1}, 
we have that $1_h<1_r$. Then Lemma \ref{lem:swapornot} for $b_r$ and $b_h$ with respect to $\prec^{opt}$, together with the optimality of 
$\prec^{opt}$,  implies that b) holds. So there is $b_{q}$ with $b_r\prec^{opt} b_{q}\prec^{opt} b_h$ such that 
$b_r\precf b_{q}$ and $1_{q}<1_h$. Now,  $b_h\prec^c b_r\precf b_{q}$ yields $b_h\prec^c b_{q}$ and thus
by Lemma \ref{lem:Q1} we should have either $b_h\precf b_{q}$ or $1_h<1_{q}$. But none of these conditions hold,
a contradiction.
\eproof

\section{Conclusion}\label{sec:conclusion}

In this paper, we proposed a new characterization for $A$-Stick graphs and deduced a new recognition algorithm for them. 
Among the three problems {\sc STICK}, {\sc STICK$_A$} and 
{\sc STICK$_{AB}$}, linear algorithms are now available for the last two. Problem  {\sc STICK} remains open. The difficulty of finding properties able to characterize Stick graphs is 
real,  in a context where very few properties of Stick graphs have been discovered. The algorithmic 
gap between  {\sc STICK$_A$} and {\sc STICK$_{AB}$}, on the one hand, and {\sc STICK}, on the other hand, seems therefore huge as of now.

Another open problem, which may be more affordable since the graph is a Stick graph by hypothesis, is {\sc MinLength Stick} 
that we proposed here and for which we provided only partial results.

\bibliographystyle{plain}
\bibliography{Stick3}

\begin{thebibliography}{10}

\bibitem{baruah2013intersection}
Arun~Kumar Baruah and Niky Baruah.
\newblock Intersection graph in traffic control problem.
\newblock {\em International Journal of Mathematics and Computer Application
  Research}, 3(01):265--270, 2013.

\bibitem{brandstadt1999graph}
Andreas Brandstadt, Van~Bang Le, and Jeremy~P. Spinrad.
\newblock {\em Graph classes: a survey}.
\newblock SIAM, 1999.

\bibitem{chaplick2018grid}
Steven Chaplick, Stefan Felsner, Udo Hoffmann, and Veit Wiechert.
\newblock Grid intersection graphs and order dimension.
\newblock {\em Order}, 35(2):363--391, 2018.

\bibitem{chaplick2019recognizing2}
Steven Chaplick, Philipp Kindermann, Andre L{\"o}ffler, Florian Thiele,
  Alexander Wolff, Alexander Zaft, and Johannes Zink.
\newblock Stick graphs with length constraints.
\newblock In {\em International Symposium on Graph Drawing and Network
  Visualization}, pages 3--17. Springer, 2019.

\bibitem{luca2018recognition2}
Felice De~Luca, Md~Iqbal Hossain, Stephen Kobourov, Anna Lubiw, and Debajyoti
  Mondal.
\newblock Recognition and drawing of {S}tick graphs.
\newblock {\em Theoretical Computer Science}, 796:22--33, 2019.

\bibitem{halldorsson2011clark}
Bjarni~V Halld{\'o}rsson, Derek Aguiar, Ryan Tarpine, and Sorin Istrail.
\newblock The clark phaseable sample size problem: long-range phasing and loss
  of heterozygosity in gwas.
\newblock {\em Journal of Computational Biology}, 18(3):323--333, 2011.

\bibitem{mckee1999topics}
T.A. McKee and F.R. McMorris.
\newblock {\em Topics in Intersection Graph Theory}.
\newblock Monographs on Discrete Mathematics and Applications. Society for
  Industrial and Applied Mathematics, 1999.

\bibitem{rusu2020stick}
Irena Rusu.
\newblock Stick graphs: examples and counter-examples.
\newblock {\em arXiv preprint arXiv:2007.10773}, 2020.

\bibitem{shrestha2011two}
Anish Man~Singh Shrestha, Asahi Takaoka, Satoshi Tayu, and Shuichi Ueno.
\newblock On two problems of nano-pla design.
\newblock {\em IEICE transactions on information and systems}, 94(1):35--41,
  2011.

\bibitem{sinden1966topology}
Frank~W Sinden.
\newblock Topology of thin film {R}{C} circuits.
\newblock {\em Bell System Technical Journal}, 45(9):1639--1662, 1966.

\end{thebibliography}
\end{document}